\documentclass[preprint,review,12pt]{elsarticle}

\usepackage{url}

\usepackage{amsmath,amssymb,amsfonts}
\usepackage[]{algorithmicx}
\usepackage{graphicx}
\usepackage{textcomp}
\usepackage{xcolor}
\usepackage{color}
\usepackage{algorithm}
\usepackage[noend]{algpseudocode}
\usepackage{booktabs,makecell} 

\newfloat{function}{htb}{}
\floatname{function}{Function}
\newfloat{procedure}{htb}{}
\floatname{procedure}{Procedure}
\usepackage[symbol]{footmisc}

\newtheorem{theorem}{Theorem}
\newtheorem{example}{Example}%
\newtheorem{definition}{Definition}
\numberwithin{equation}{section}

\newtheorem{lemma}[theorem]{Lemma}
\usepackage{wrapfig}
\usepackage{pdflscape}
\usepackage{rotating}
\usepackage{epstopdf}

\usepackage{amssymb}

\journal{Knowledge-Based Systems}

\begin{document}

\begin{frontmatter}



\title{An Efficient Approach for Discovering Graph Entity Dependencies (GEDs)}


\author{Dehua Liu\fnref{ci}}
\affiliation[ci]{%
    organization={College of Informatics},
    addressline={Huazhong Agric. University},
    city={Wuhan},
    state={Hubei},
    country={China}
}

\author{Selasi Kwashie\fnref{cd}}
\affiliation[cd]{%
    organization={AI \& Cyber Futures Institute},
    addressline={Charles Sturt University},
    city={Bathurst},
    state={NSW},
    country={Australia}
}

\author{Yidi Zhang\fnref{ci}}
\author{Guangtong Zhou\fnref{ci}}

\author{Michael Bewong\fnref{scm}}
\affiliation[scm]{%
    organization={School of Computing \& Mathematics},
    addressline={Charles Sturt University},
    city={Wagga Wagga},
    state={NSW},
    country={Australia}
}

\author{Xiaoying Wu\fnref{sc}}
\affiliation[sc]{%
    organization={School of Computer},
    addressline={Wuhan University},
    city={Wuhan},
    state={Hubei},
    country={China}
}
\author{Xi Guo\fnref{ci}}

\author{Keqing He\fnref{sc}}

\author{Zaiwen Feng\corref{ca}\fnref{ci}}
\ead[]{Zaiwen.Feng@mail.hzau.edu.cn}
\cortext[ca]{Corresponding author}

\begin{abstract}
Graph entity dependencies (GEDs) are novel graph constraints, unifying keys and functional
dependencies, for property graphs. They have been found useful in many real-world data quality 
and data management tasks, including fact checking on social media networks and entity resolution. 
In this paper, we study the discovery problem of GEDs -- finding a minimal cover of valid GEDs 
in a given graph data. We formalise the problem, and propose an effective and efficient approach
to overcome major bottlenecks in GED discovery. In particular, we leverage existing graph 
partitioning algorithms to enable fast GED-scope discovery, and employ effective pruning
strategies over the prohibitively large space of candidate dependencies.  
Furthermore, we define an interestingness measure for GEDs based on the minimum description 
length principle, to score and rank the mined cover set of GEDs. 
Finally, we demonstrate the scalability and effectiveness of our GED discovery approach through
extensive experiments on real-world benchmark graph data sets; and present the usefulness of
the discovered rules in different downstream data quality management applications. 
\end{abstract}


\begin{highlights}
\item A study of the discovery problem of Graph Entity Dependencies (GEDs).
\item A new and efficient approach for the discovery of GEDs in
property graphs.
\item A minimum description length inspired definition of interestingness of
GEDs to rank discovered rules.
\item A thorough empirical evaluation of the proposed technique, with examples of
useful rules (mined) that are relevant in data quality/management applications.
\end{highlights}

\begin{keyword}
Graph entity dependencies \sep Graph entity dependency discovery \sep Efficient algorithm
\sep Data Dependencies



\end{keyword}

\end{frontmatter}

\section{Introduction}\label{sec:intro}
In recent years, integrity constraints (e.g., {\em keys}~\cite{gkeys} and {\em functional 
dependencies} (FDs)~\cite{gfd}) have been proposed for property graphs to enable
the specification of various data semantics, and tackling of graph data quality 
and management problems. Graph entity dependencies (GEDs)~\cite{b0,ged} are new 
fundamental constraints unifying keys and FDs for property graphs. 

A GED $\varphi$ over a graph $G$ is a pair, $\varphi = (Q[\bar{u}], X\to Y)$, specifying the
dependency $X\to Y$ over {\em homomorphic matches} of the graph pattern $Q[\bar{u}]$ in $G$. 
Intuitively, since graphs are schemaless, the graph pattern $Q[\bar{u}]$ identifies
which the set of entities in $G$ on which the dependency $X\to Y$ should hold. 
In general, GEDs are capable of expressing and encoding the semantics of relational
FDs and conditional FDs (CFDs), as well as subsume other graph integrity constraints
(e.g., GKeys~\cite{gkeys} and  GFDs~\cite{gfd}). 

GEDs have numerous real-world data quality and data management applications (cf.~\cite{b0,ged} 
for more details). For example, they have been extended and used for: fact checking in 
social media networks~\cite{fact_check}, entity resolution in graphs and relations~\cite{ontology,her, 
b1}, consistency checking~\cite{incons}, knowledge graph generation~\cite{kb-feng}, amongst others.  

In this work, motivated by the expressivity and usefulness of GEDs in many downstream 
applications, we study the problem of automatically learning GEDs from a given property 
graph. To the best of our knowledge, at the moment, no discovery algorithm exists in 
the literature for GEDs, albeit discovery solutions have been proposed for {\em graph 
keys} and GFDs in~\cite{gkeyminer} and \cite{b3} respectively. 
However, the GFD mining solutions in~\cite{b3} mine a special class of GEDs (without any id-literals, using isomorphic matching semantics). Further, 
\cite{b3,gkeyminer} considers the discovery of key constraints in RDFs.
Thus, the need for effective and efficient techniques for mining rules that capture the
full GED semantics still exists. 

Discovering data dependencies is an active and long-standing research problem within the database
and data mining communities. Indeed, the problem is well-studied for the relational
data setting, with volumes of contributions for functional dependencies (FDs)~\cite{profiling,
binder,b19,b5} and its numerous extensions (e.g., conditional FDs~\cite{review,mineCFD,FindCFD}, distance-based FDs~\cite{mfd, relaxed, scamdd, ardd, cdd}, etc).

The general goal in data dependency mining is to find an irreducible set (aka. {\em minimal cover set}) 
of valid dependencies (devoid of all implications and redundancies) that hold over the given input data. This is a challenging and often intractable problem for most dependencies, and GEDs are not an exception. 
In fact, the problem is inherently more challenging for GEDs than it is for other graph constraints
(e.g. GFDs and GKeys) and traditional relational dependencies. The reason is three-fold:
a) the presence of graph pattern as topological constraints (unseen in relational dependencies);
b) the attribute sets of GED dependency have three possible literals (e.g., GFDs have 2, GKeys have 1, CFDs have 2);
c) the implication and validation analyses---{\em key tasks in the discovery process}---of GEDs are
shown NP-complete and co-NP-complete respectively (see~\cite{ged} for details). 

This paper proposes an efficient and effective GED mining solution. 
We observe that, two major efficiency bottlenecks in mining GEDs are on 
pattern discovery in large graphs, traversal of the prohibitively large search space 
of candidate dependencies. 
Thus, we leverage existing graph partitioning algorithms to surmount the first challenge,
and employ several effective pruning rules and strategies to resolve the second. 

We summarise our main contributions as follows. 
1) We formalise the GED discovery problem (in Section~\ref{sec:pro}). We extend the notions of
{\em trivial, reduced}, and {\em minimal} dependencies for GEDs; and introduce a formal cover set
finding problem for GEDs.
2) We develop a new two-phase approach for efficiently finding GEDs in large graphs
(Setion~\ref{sec:sol}). The developed 
solution leverages existing graph partitioning algorithms to mine graph patterns quickly, and 
find their matches. Further, we design a comprehensive attribute (and entity) dependency discovery
architecture with effective candidate generation, search space pruning and validation strategies. 
Moreover, we introduce a new interestingness score for GEDs, for ranking the discovered rules. 
3) We perform extensive experiments on real-world graphs to show that GED discovery can
be feasible in large graphs, and demonstrate the effectiveness, scalability and 
efficiency of our proposed approach (Section~\ref{sec:exp}).

\section{Preliminaries}\label{sec:pre}

This section presents preliminary concepts and definitions.
We use $\mathbf{A, L, C}$ to be denote universal sets of {\em attributes,
labels} and {\em constants} respectively. 
The definitions of {\it graph, graph pattern}, and {\it matches} follow 
those in~\cite{b0,ged}.

{\bfseries Graph.}
We consider a directed property graph $G=(V,E,L,F_{A})$, where: (1) $V$ is a finite set 
of nodes; (2) $E$ is a finite set of edges, given by $E \subseteq V\times \mathbf{L}\times V$, 
in which $(v,l,v')$ is an edge from node $v$ to node $v'$ with label $l \in \mathbf{L}$;  
(3) each node $v \in V$ has a special attribute $id$ denoting its identity, and a label $L(v)$
drawn from $\mathbf{L}$; (4) every node $v$, has an associated list
$F_A(v)=[(A_{1},c_{1}),...,(A_{n},c_{n})]$ of attribute-value pairs, where 
$c_{i} \in \mathbf{C}$ is a constant, $A_{i} \in \mathbf{A}$ 
is an attribute of $v$, written as $v.A_{i}=c_{i}$, and $A_{i} \neq A_{j}$ if $i \neq j$.

{\bfseries Graph Pattern.}
A graph pattern, denoted by $Q[\bar{u}]$, is a directed graph $Q[\bar{u}]=(V_{Q}, E_{Q}, L_{Q})$,
where: (a) $V_{Q}$ and $E_{Q}$ represent the set of pattern nodes and pattern edges respectively;
(b) $L_{Q}$ is a label function that assigns a label to each node $v\in V_Q$ and each edge 
$e\in E_Q$; and (c) $\bar{u}$ is all the nodes, called (pattern) variables in $V_{Q}$.
All labels are drawn from $\mathbf{L}$, including the wildcard ``*" as a special label.
Two labels $l,l'\in \mathbf{L}$ are said to {\em match}, denoted $l\asymp l'$ iff: 
(a) $l=l'$; or (b) either $l$ or $l'$ is ``*". 

A {\bf match} of a graph pattern $Q[\bar{u}]$ in a graph $G$ is a homomorphism $h$ 
from $Q$ to $G$ such that: (a) for each node $v\in V_Q$, $L_Q(v) \asymp L(h(v))$;
and (b) each edge $e = (v,l,v') \in E_Q$, there exists an edge $e'=(h(v), l', h(v'))$
in $G$, such that $l\asymp l'$.
We denote the list of all matches of $Q[\bar{u}]$ in $G$ by $H(\bar{u})$.
An example of a graph, graph patterns and their matches are presented below in 
Example~\ref{ex:graph}. 


\begin{figure}[t]
\centering
\includegraphics[width=\linewidth]{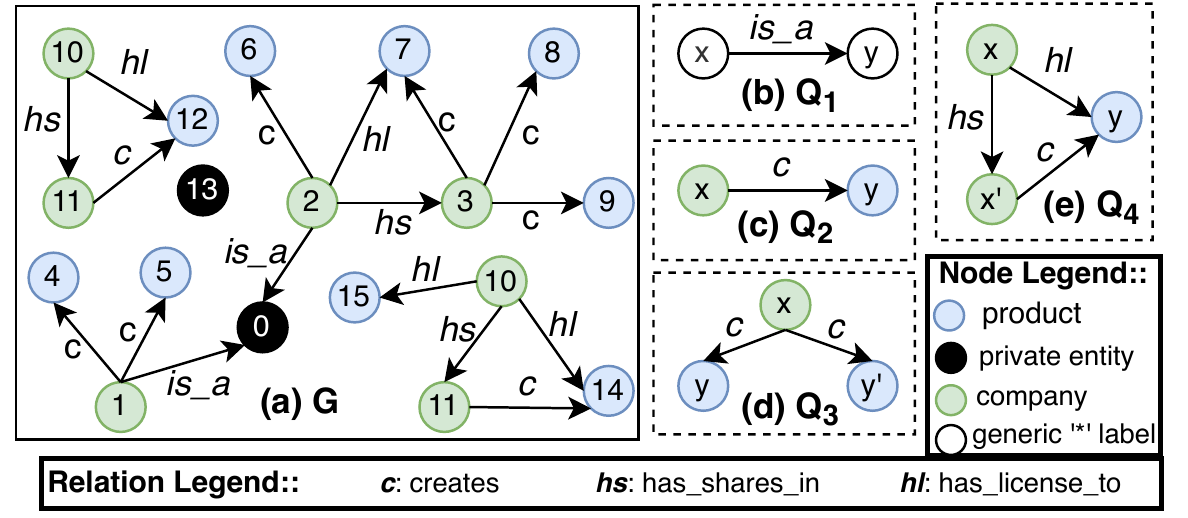}
\caption{Example: (a) graph; (b)--(d) graph patterns}
\label{fig:GED}
\end{figure}

\begin{example}[Graph, Graph Pattern \& Matching]\label{ex:graph}
Figure~\ref{fig:GED}(a) depicts a simple graph; and  (b)--(e) are examples of four graph 
patterns. 
$Q_{1}[x,y]$ describes an ``\verb|is a|" relationship between two generic (``*" labelled)
nodes $x$ and $y$. The list of matches of this pattern in the example graph is 
$H_1(x,y)=[\{1, 0\}, \{2, 0\}]$.
$Q_2[x,y]$ depicts a \verb|company| node $x$ with a \verb|create| relation with a \verb|product| 
node $y$, its matches in $G: H_2(x,y)=[\{1, 4\} \{1, 5\} \{2, 6\}, \{3, 7\}, \{3, 8\}$, $\{3,9\},
\{11, 12\}, \{11, 14\}]$.
$Q_3[x,y,y']$ specifies a \verb|company| node $x$ with a \verb|create| relation with 
two \verb|product| nodes $y,y'$. Thus, matches in $G$ are $H_3(x,y,y') = [\{1,4,5\}, \{3,7,8\}, \{3,7,9\}, \{3,8,9\}]$.
$Q_4[x,x',y]$ describes a \verb|company| $x$ with a \verb|has shares in| relation with 
another \verb|company| $x'$, and a \verb|has license to| relation with a \verb|product| $y$
created by $x'$. $H_4(x,x',y) =[\{2,4,7\}, \{10, 11, 12\}, \{10$, $11, 14\}]$. 
\end{example}

{\bfseries Graph Entity Dependencies.}
A graph entity dependency (GED)~\cite{b0} $\varphi$ is a pair $(Q[\bar{u}],X \to Y)$, 
where $Q[\bar{u}]$ is a graph pattern, and $X,Y$ are two (possibly empty) sets of {\em literals}
in $\bar{u}$. 
A literal $w$ of $\bar{u}$ is one of the following three constraints:
(a) $x.A = c$, (b) $x.A = y.A'$, (c) $x.id =y.id$, where
$x, y\in \bar{u}$, are pattern variables, $A,A' \in \mathbf{A}$, are non-$id$ attributes,
and $c\in \mathbf{C}$ is a constant. 
$Q[\bar{u}]$, and $X\to Y$ are referred to as the {\em pattern/scope} and {\em dependency} of 
$\varphi$ respectively. 

Given a GED $\varphi=(Q[\bar{u}], X\to Y)$, a match $h(\bar{u})$ of $Q[\bar{u}]$ in $G$ 
{\em satisfies} a literal $w$ of $\bar{u}$, denoted by $h(\bar{u}) \models w$, if:
(a) when $w$ is $x.A = c$, then the attribute $h(x).A$ exists, and $h(x).A = c$; 
(b) when $w$ is $x.A = y.A'$, then the attributes $h(x).A$ and $h(y).A'$ exist
and have the same value; (c) when $w$ is $x.id = y.id$, then $h(x).id = h(y).id$. 

A match $h(\bar{u})$ satisfies a set $X$ of literals if $h(\bar{u})$ satisfies every literal 
$w \in X$, (similarly defined for $h(\bar{u}) \models Y$).
We write $h(\bar{u}) \models X \to Y$ if $h(\bar{x}) \models X$ implies $h(\bar{x}) \models Y$.

A graph $G$ satisfies GED $\varphi$, denoted by $G \models \varphi$, if for all matches $h(\bar{u})$ of $Q[\bar{u}]$ in $G$, $h(\bar{x}) \models X \to Y$. 
A graph, $G$, satisfies a set $\Sigma$ of GEDs, denoted by $G \models \Sigma$, if for all 
$\varphi \in \Sigma$, $G \models \varphi$.

We illustrate the semantics of GEDs via the sample graph and graph patterns in 
Figure~\ref{fig:GED} below. 

\begin{example}[GEDs]
We define exemplar GEDs over the sample graph in Figure~\ref{ex:graph}(a), using the graph 
patterns in Figure~\ref{ex:graph}(b)--(e).
1) $\varphi_1: (Q_1[x,y], y.A=y.A\to x.A=y.A)$ -- this GED states that for any match $h(x,y)$ 
of the pattern $Q_1[x,y]$ in $G$ (i.e., $x$ \verb|is_a| $y$), if the node $h(y)$ has property
$A$, then $h(x)$ must have same values as $h(y)$ on $A$. 
2) $\varphi_2: (Q_2[x,y], \emptyset \to y.creator=x.name)$ -- for every match $h(x,y)$ of $Q_2$ in $G$
(i.e., \verb|company| $h(x)$ \verb|create| \verb|product| $h(y)$), then $h(y.creator)$ and 
$h(x.name)$ must have the same value. 
3) $\varphi_3: (Q_3[x,y,y'], \emptyset\to y.creator=y'.creator)$ -- this states for any match 
$h(x,y,y')$ in $G$ (i.e., the \verb|company| $h(x)$ \verb|create|
two products $h(y,y')$), then $y,y'$ must have same value on their property/attribute $creator$.
4) $\varphi_4: (Q_4[x,x',y], x'.country=$ \verb|USA|$ \to x.country=$ \verb|USA|) -- for every match $h(x,x',y)$ of $Q_4$ in $G$, if $h(x'.country)$ is \verb|USA|, then $h(x.country)$ must also be \verb|USA|. Note that, $\varphi_4$ over $G$ restricts share/license ownership in any USA company to only USA companies. 
\end{example}

\section{Problem Formulation}\label{sec:pro}
Given a property graph $G$, the general GED discovery problem is to find a set $\Sigma$
of GEDs such that $G\models \Sigma$. However, like other dependency discovery problems,
such a set can be large and littered with redundancies (cf.~\cite{b3,b1,b19}). 
Thus, in line with~\cite{b3}, we are interested in finding a {\em cover} set of {\em non-trivial} 
and {\em non-redundant} dependencies over {\em persistent} graph patterns. 
In the following, we formally introduce the GEDs of interest, and present our 
problem definition.

{\bfseries Persistent Graph Patterns.} Graph patterns can be seen as ``loose schemas" 
in GEDs~\cite{ged}. It is therefore crucial to find persistent graph patterns in the 
input graph data for GED discovery. However, counting the frequency of sub-graphs is 
a non-trivial task, and can be intractable depending on the adopted count-metric 
(e.g., harmful overlap~\cite{ho}, and maximum independent sets~\cite{mis}). 

We adopt the {\em minimum image based support} (MNI)~\cite{mni} metric 
to count the occurrence of a graph pattern in a given graph, due to its anti-monotone
and tractability properties. 

Let $I=\{i_1, \cdots, i_m\}$ be the set of isomorphisms of a pattern $Q[\bar{u}]$ to
a graph $G$. Let $M(v) = \{i_1(v), \cdots, i_m(v)\}$ be the set containing the (distinct) 
nodes in $G$ whose functions $i_1,\cdots, i_m$ map a node $v\in V$. 
The MNI of $Q$ in $G$, denoted by $mni(Q, G)$, is defined as:
\begin{align}\label{eq:mni}
mni(Q, G) = min\{x\mid x =|M(v)|,\mbox{ } \forall \mbox{ }v \in  V \}. 
\end{align}

We say a graph pattern $Q[\bar{u}]$ is {\em persistent} (i.e., {\em frequent}) in a graph $G$,
if $mni(Q, G) \geq \tau$, where $\tau$ is a user-specified minimum MNI threshold. 

{\bf Representative Graph Patterns.} For many real-world graphs, it is prohibitively 
inefficient to mine frequent graph patterns with commodity computational resources largely 
due to their sizes (cf.~\cite{surv_sg_mining, surv_gp_mining}). 
To alleviate this performance bottleneck, in our pipeline, we consider graph 
patterns that are frequent within densely-connected communities within the input graph. 
We term such graph patterns {\it representative}, and adopt the Constant Potts Model 
(CPM)~\cite{cpm} as the quality function for communities in a graph. 

Formally, the CPM of a community $c$ is given by
\begin{align}\label{eq:cpm}
    \mathcal{H} = \sum_c\left[ e_c - \gamma \binom{n_c}{2} \right]
\end{align}
where $e_c$ and $n_c$ are the number of edges and nodes in $c$ respectively; and 
$\gamma>0$ is a (user-specified) resolution parameter---higher $\gamma$ leads 
to more communities, and vice versa. Intuitively, the resolution parameter, $\gamma$, 
constrains the intra- and inter-community densities to be no less than 
and no more than the value of $\gamma$ respectively. 

Thus, we say a graph pattern $Q[\bar{u}]$ is {\it representative} or $(\gamma, \tau)$-frequent
if it is $\tau$-frequent within at least one $\gamma$-dense community $c\in G$. 

{\bfseries Trivial GEDs.} We say a GED $\varphi: (Q[\bar{u}], X\to Y)$ is {\em trivial}
if: (a) the set of literals in $X$ cannot be satisfied (i.e., $X$ evaluates to {\tt false}); 
or (b) $Y$ is derivable from $X$ (i.e., $\forall \mbox{ }w\in Y$, $w$ can be derived from $X$
by transitivity of the equality operator). We consider non-trivial GEDs.

{\bfseries Reduced GEDs.} Given two patterns $Q[\bar{u}]=(V_Q,E_Q,L_Q), Q'[\bar{u'}]=
(V'_{Q'},E'_{Q'}, L'_{Q'})$, $Q[\bar{u}]$ is said to {\em reduce} $Q'[\bar{u'}]$, denoted 
as $Q\ll Q'$ if: (a) $V_Q\subseteq V'_{Q'}$, $E_Q\subseteq E'_{Q'}$; or (b) $L_Q$ upgrades
some labels in $L'_{Q'}$ to wildcards. 
That is, $Q$ is a less restrictive topological constraint than $Q'$.

Given two GEDs, $\varphi = (Q[\bar{u}], X\to w)$ and $\varphi' = (Q'[\bar{u'}], X'\to w')$.  
$\varphi$ is said to {\em reduce} $\varphi'$, denoted by $\varphi\ll \varphi'$,
if: (a) $Q\ll Q'$; and (b) $X\subseteq X'$ and $w=w'$.

Thus, we say a GED $\varphi$ is {\bf reduced} in a graph $G$ if: (a) $G\models \varphi$;
and for any $\varphi'$ such that $\varphi'\ll \varphi$, $G\not\models\varphi'$.
We say a GED $\varphi$ is {\bf minimal} if it is both non-trivial and reduced. 

{\bfseries Cover of GEDs.} Given a set $\Sigma$ of GEDs such that $G\models \Sigma$. 
$\Sigma$ is {\em minimal} iff: for all $\varphi\in \Sigma$, we have
$\Sigma\not\equiv\Sigma\setminus\varphi$. That is, $\Sigma$ contains no redundant GED. 
We say $\Sigma_c$ is a cover of $\Sigma$ on $G$ if: (a) $\Sigma_c\subseteq\Sigma$;
(b) $\Sigma_c\equiv\Sigma$; (c) all $\varphi \in \Sigma_c$ are minimal; and (d) $\Sigma_c$
is minimal. 


We study the following GED discovery problem. 

\begin{definition}[Discovery of GEDs]
Given a property graph, $G$, and user-specified resolution parameter and MNI thresholds 
$\gamma, \tau$:
find a cover set $\Sigma_c$ of all valid minimal GEDs that hold over $(\gamma, \tau)$-frequent 
graph patterns in $G$.
\end{definition}

\begin{figure}[!h]
\centering
\includegraphics[width=\linewidth]{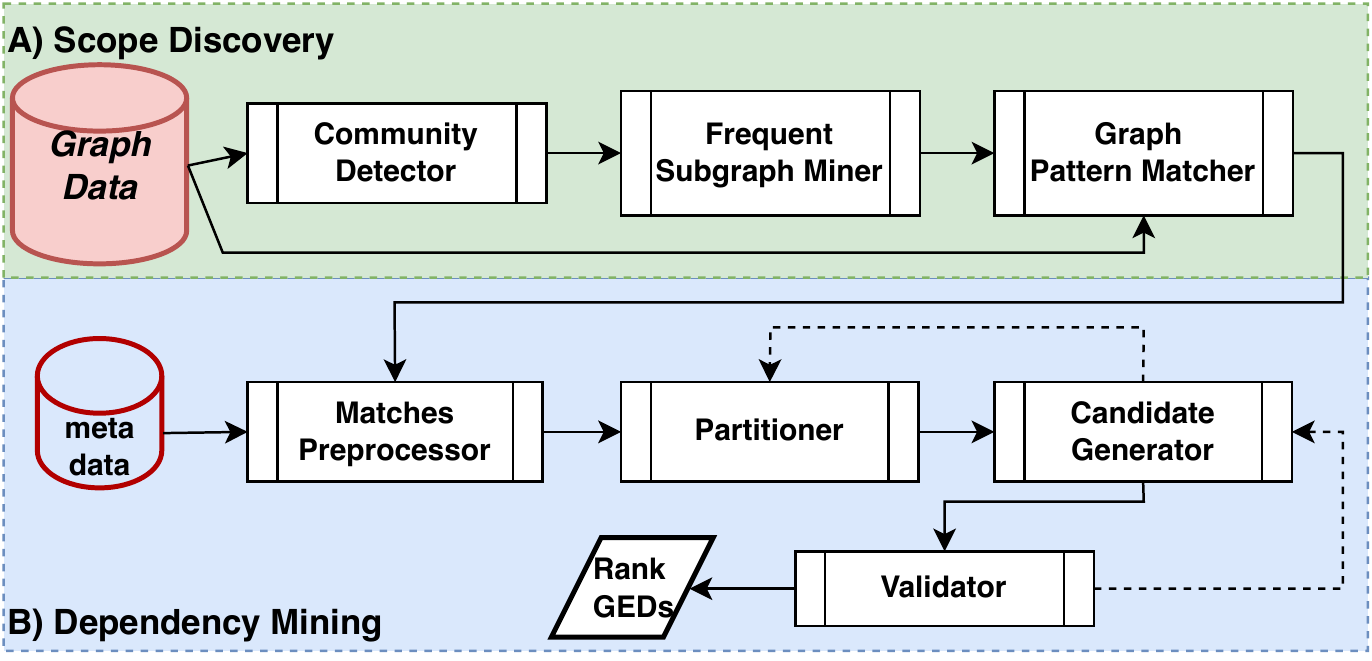}
\caption{The Overall GED Discovery Pipeline.}
\label{fig:pipeLine}
\end{figure}

\section{The Proposed Discovery Approach}\label{sec:sol}

This section introduces a new and efficient GED discovery approach. 
Figure~\ref{fig:pipeLine} presents the overall pipeline of our solution, 
consisting of two main components for: (a) finding {\em representative} and 
{\it reduced} graph patterns (i.e., ``scopes'') and their (homomorphic) matches in the graph; 
and (b) finding {\em minimal} attribute (entity) dependencies that holds over 
the graph patterns. A snippet of the pseudo-code for the proposed GED discovery process 
is presented Algorithm~\ref{alg:gedminer}, requiring two main procedures for the 
above-mentioned tasks (cf.~lines~\ref{al-minescope} and~\ref{al-minedep}--\ref{al-minedep-end} of Algorithm~\ref{alg:gedminer} respectively). 
The required input to the algorithm includes: a
property graph, $G$; a user-specified resolution parameter $\gamma$ and minimum MNI 
threshold $\tau$.

\begin{algorithm}[t]
\caption{DisGEDs}
\label{alg:gedminer}
\begin{algorithmic}[1] 
\Require A graph, $G$; CPM resolution-parameter, $\gamma$; MNI threshold, $\tau$. 
\Ensure A cover set $\Sigma_c$ of GEDs (with $\tau$-frequent graph patterns) in $G$
\State{$\Sigma_c:=\emptyset; \Sigma:=\emptyset; \mathcal{S}:=\emptyset$}
\State{$\mathcal{S}\leftarrow mineScopes(G,\gamma, \tau)$}\Comment{find graph patterns \& their matches}\label{al-minescope}
\For{{\bf each} $(Q_i, H(Q_i,G)) \in \mathcal{S}$}\Comment{mine minimal GEDs}\label{al-minedep}
\State{$\Sigma_i \leftarrow mineDependencies(Q_i,H(Q_i, G))$}
\State{$\Sigma \leftarrow \Sigma\cup\Sigma_i$}
\EndFor \label{al-minedep-end}
\State{$\Sigma_c\leftarrow findCover(\Sigma)$}\Comment{remove all transitively implied GEDs}\label{al-prune}
\State{\Return $\Sigma_c$}
\end{algorithmic}
\end{algorithm}

\subsection{Graph Pattern Discovery \& Matching}\label{ssec:fsm}
The first task in the discovery process is finding representative graph patterns, 
and their matches in the given graph. We present a three-phase process for the task, 
viz.: (i) detection of dense communities within the input graph; (ii) mining frequent 
graph patterns in communities; and (iii) finding matches of the discovered (and reduced) 
graph patterns in the input graph. 

Procedure~\ref{alg:patminer}, mineScopes($G, \gamma, \tau$), presents the pseudo-code for 
this task (with workflow depicted in Figure~\ref{fig:pipeLine}~A). 

\subsubsection{Dense communities detection}
The first step in our approach is straightforward: division of the input 
\begin{wrapfigure}[9]{r}{0.45\textwidth}
    \centering
    \includegraphics{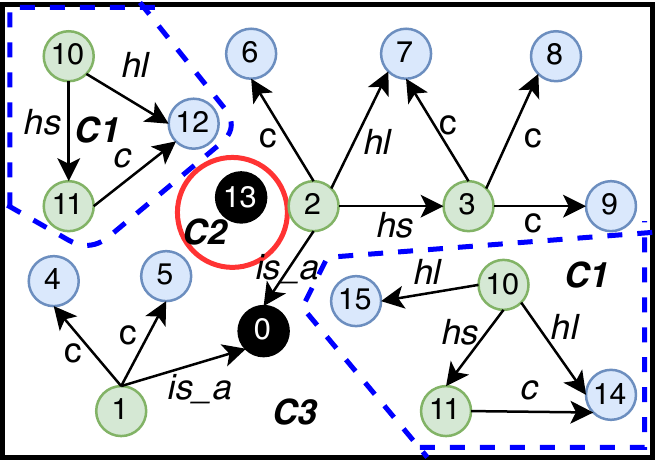}
    \caption{\(0.01\)-dense communities in example graph $G: C1, C2, C3$}
    \label{fig:splitting}
\end{wrapfigure}
graph $G$ into 
multiple dense communities, $\mathcal{C} =\{c_1, \cdots, c_k$\}. In particular, we employ the
quality function in Equation~\ref{eq:cpm} to ensure that the density of
each community $c_i\in \mathcal{C}, (i\in [1,k])$ is no less than $\gamma$. 
In general, any efficient graph clustering algorithm can be employed for this task;
and we adopt the {\tt Leiden} graph clustering algorithm~\cite{b11} with the CPM quality
function to partition the input graph into $\gamma$-dense communities (cf. line~\ref{fl-leiden}
in Procedure~\ref{alg:patminer}).  

As an example, consider our sample graph in Figure~\ref{fig:GED}~(a).
For a resolution value of $\gamma=0.01$, three dense communities are produced as show in
Figure~\ref{fig:splitting}. This includes: $c_1$, portion of the graph enclosed by
the dotted blue lines; $c_2$, a single node \(13\); and $c_3$, the remaining graph.


\begin{procedure}
\caption{mineScopes($G, \gamma, \tau$)}
\label{alg:patminer}
\begin{algorithmic}[1] 
\Require Graph, $G$; CPM resolution-parameter, $\gamma$; MNI threshold, $\tau$
\Ensure Set $\mathcal{S} =\{(Q_1, H(Q_1, G)), \cdots\}$ of reduced patterns $Q_i$ and their homomorphic matches $H(Q_i, G)$ in $G$
\State{$\mathcal{C}:=\emptyset; \mathcal{Q}:=\emptyset; \mathcal{S}:=\emptyset; G'=\emptyset$}
\State{$\mathcal{C}\leftarrow {\tt Leiden}(G, \gamma)$}\Comment{mine dense communities}\label{fl-leiden}
\For{$c_i\in \mathcal{C}$}\Comment{mine frequent sup-graph patterns}\label{fl-mine-gp}
\State{$\mathcal{Q}_i \leftarrow {\tt GraMi}(c_i,\tau)$}
\State{$\mathcal{Q}\leftarrow \mathcal{Q}\cup\mathcal{Q}_i$}
\EndFor\label{fl-mine-gpe}
\State{$\mathcal{Q}'\leftarrow reduceGraphPatterns(\mathcal{Q})$}\label{fl-reduce}
\For{$Q \in \mathcal{Q}'$}\Comment{find homomorphic matches}\label{fl-match}
\State{$G' \leftarrow {\tt filter}(G, Q)$}\Comment{filter $G$ w.r.t. edges/nodes in $Q$}
\State{$H(Q,G) \leftarrow {\tt match}(Q, G')$}
\State{$\mathcal{S} \leftarrow \mathcal{S}\cup (Q, H(Q, G))$}
\EndFor\label{fl-match-end}
\State{\Return $\mathcal{S}$}
\end{algorithmic}
\end{procedure}

\subsubsection{Representative Graph Patterns Mining}
Next, we take the discovered $\gamma$-dense communities, $\mathcal{C}=\{c_1, \cdots, c_k\}$ as input, 
and return a set of {\em frequent} and reduced graph patterns, using 
isomorphic matching and the MNI metric (in Equation~\ref{eq:mni}) to count patterns 
occurrences in the communities (cf. lines~\ref{fl-mine-gp}--\ref{fl-reduce} of Procedure~\ref{alg:patminer}).


We adapt the {\tt GraMi} graph patterns mining algorithm~\cite{b7} to find the set, $\mathcal{Q}$, of all 
$\tau$-frequent patterns across $c_1, \cdots, c_k$. As will be expected, $\mathcal{Q}$ may contain redundant patterns. 
Therefore, we designed a simple and effective algorithm, Function~\ref{alg:A}, to prune all redundancies 
in $\mathcal{Q}$, and return a set $\mathcal{Q}'$ of representative and reduced patterns in $G$ 
(i.e., there does not exist any pair of patterns $Q,Q'\in \mathcal{Q}'$ such that $Q\ll Q'$ or $Q'\ll Q$).

\begin{function}
\caption{$reduceGraphPatterns(\mathcal{Q})$}
\label{alg:A}
\begin{algorithmic}[1] 
\State{$n\leftarrow |\mathcal{Q}|; \mathcal{Q}':= \emptyset;
\Vec{T} \leftarrow [false,false,...,false]
$}\label{fl-ini}
\State{$Sort(\mathcal{Q})$}\Comment{in descending order of pattern sizes}\label{fl-sort}
\For{$i \leftarrow 0$ to $n-1$}\label{fl-tag}
\If{$\Vec{T}[i] == false$}
\For{$j \leftarrow i + 1$ to $n-1$}
\If{$\Vec{T}[j] == false$}
\If{${\tt count}({\tt VF2U}(\mathcal{Q}(i),\mathcal{Q}(j))) > 0$
}
\State{$\Vec{T}[j] \leftarrow true$}\label{fl-tag-true}
\EndIf
\EndIf
\EndFor
\State{$\mathcal{Q}' \leftarrow \mathcal{Q}' \cup \mathcal{Q}(i)$}
\EndIf
\EndFor\label{fl-tag-end}
\State{\Return $\mathcal{Q}'$}
\end{algorithmic}
\end{function}


The input to Function~\ref{alg:A} is a set $\mathcal{Q}$ of $\tau$-frequent graph patterns 
in $\gamma$-dense communities (ie., $(\gamma, \tau)$-frequent or representative graph patterns);
and the output is a set $\mathcal{Q}'$ of graph patterns without any redundancies. 

Let $n$ be the size, $|\mathcal{Q}|$, of $\mathcal{Q}$; and $\Vec{T}$ be an $n$-length 
vector where each location $\Vec{T}[i]$ at index $i \in [0,n-1]$ holds a Boolean value. 
Each entry $\Vec{T}[i]$ is associated with a graph pattern $Q_i\in \mathcal{Q}$, and denotes
whether or not the associated graph pattern is redundant.

The variables, $n, \mathcal{Q}'$ and $\Vec{T}$ are initialised in line~\ref{fl-ini} of 
Function~\ref{alg:A}. The graph patterns in $\mathcal{Q}$ are then sorted from largest 
to smallest by the number of edges in every graph pattern (in line~\ref{fl-sort}).
Thereafter, we evaluate 
    \begin{wrapfigure}[20]{r}{0.5\textwidth}
    \vspace{-12pt}
    \centering
    \includegraphics[width=0.5\textwidth]{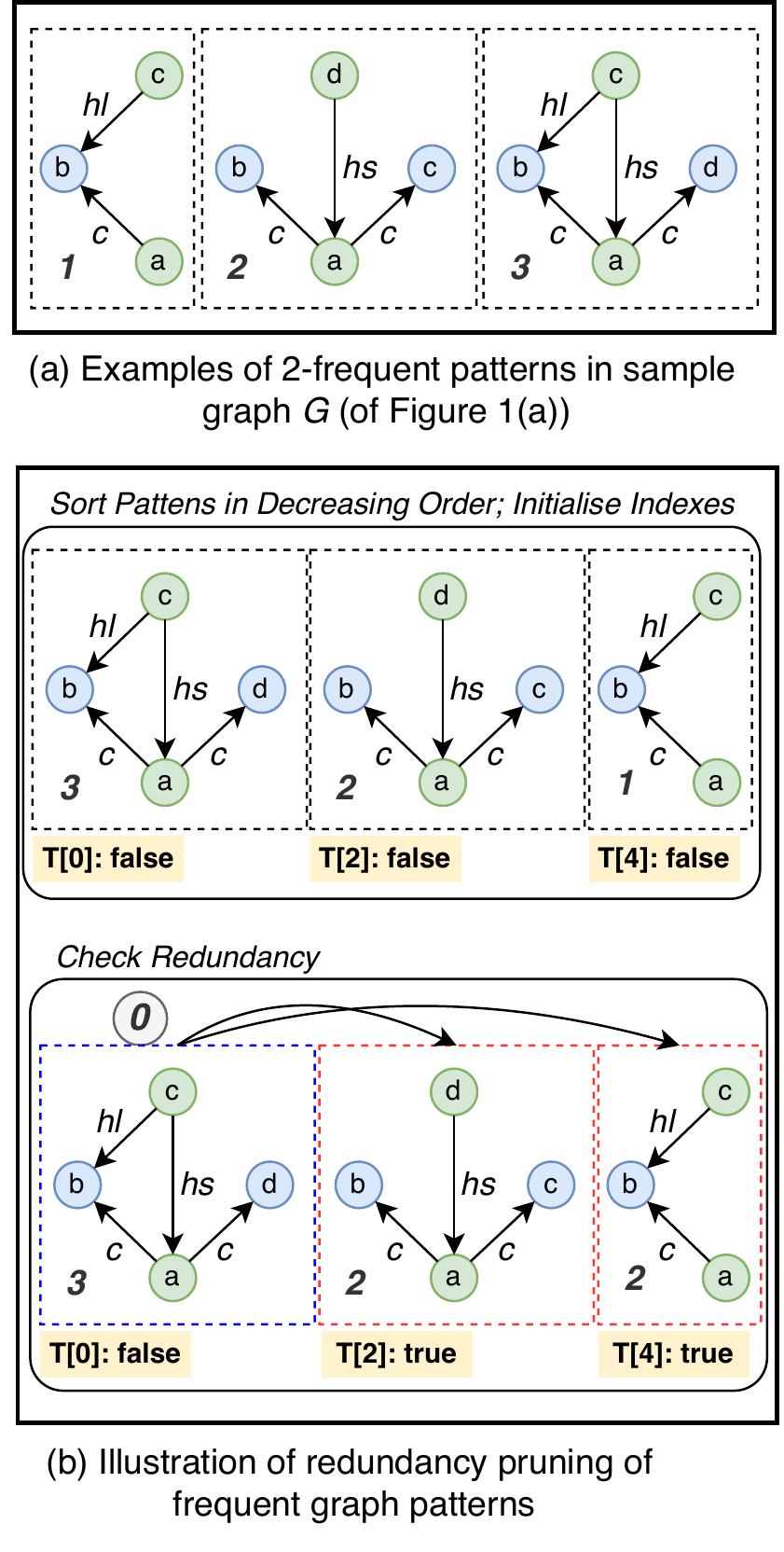}
    \vspace{-25pt}
    \caption{Frequent subgraph pruning}
    \label{fig:fsm-p}
    \end{wrapfigure}
pairs of graph patterns for redundancy (in lines~\ref{fl-tag}--\ref{fl-tag-end}). In particular, the functions {\tt VF2U} and {\tt count} are used to check isomorphic matches 
between patterns at index $i,j$ of $\mathcal{Q}$, and compute the number of isomorphic
matches returned respectively. 
We extended an efficient implementation\footnote{https://github.com/MorvanHua/VF2} of the
VF2 algorithm\cite{b18} through the relaxation of some criteria as {\tt VF2U}. The main
difference being, in {\tt VF2U}, instead of {\it induced subgraphs}, the query graph pattern 
simply queries subgraphs in the target graph pattern which are isomorphic to it. 

For the pairwise comparison of the graph patterns, the pattern at index $j$ is 
considered as the query pattern whiles that at index $i$ is the target pattern. 
The algorithm iterates from the largest pattern in $\mathcal{Q}$ (line~\ref{fl-tag}). 
In the round of iteration $i$, the largest pattern indexed $i$ is used as the $\mathcal{Q}(i)$ 
if it is tagged with $false$. As $\mathcal{Q}(j)$, every pattern indexed from $i+1$ to $n-1$ 
is checked to observe if it is isomorphic to $\mathcal{Q}(i)$. {\tt VF2U} returns the 
isomorphisms from $\mathcal{Q}(j)$ to $\mathcal{Q}(i)$, and {\tt count} returns the size of 
the resulting set. 

During every iteration, the pattern indexed $j$ ($j > i$) is tagged as $true$ (i.e., 
line~\ref{fl-tag-true}) if the $\mathcal{Q}(j)$ is isomorphic to the $\mathcal{Q}(i)$, and 
the $\mathcal{Q}(j)$ will not be a target graph in the following iterations. After all the patterns 
indexed from $i+1$ to $n-1$ are checked, the $\mathcal{Q}(i)$ is added into $\mathcal{Q}'$ and the 
round of iteration $i$ ends. Finally, when all graph patterns have been processed, the set 
$\mathcal{Q}'$ which stores all graph patterns without redundancies is returned as the output.

We present a brief illustration of the reduction process
in Example~\ref{ex:fsm}.

\begin{example}[Representative Patterns]\label{ex:fsm}
    Given the \(0.01\)-dense three communities discovered in Figure~\ref{fig:splitting}, 
    let the minimum support be $\tau=2$ for mining frequent subgraphs in the 
    respective communities. Figure~\ref{fig:fsm-p}~(a) shows a list of three (3) frequent 
    but redundant subgraphs. We illustrate the pruning strategy as follows, using the
    diagram in Figure~\ref{fig:fsm-p}~(b).

    First, we sort the subgraphs in increasing order (of edges) and initialise the
    corresponding index $i$ in the vector $\Vec{T}[i]$ as {\tt false}. Next, we
    iterate over the vector, from the largest subgraph at $i$ and the next at
    $j=i+1$, to check the containment of the pattern at $j$ in the pattern at $i$.
    We set the index at $\Vec{T}[j]$ to {\tt true} if it contained in the patern at $i$.

    Finally, we return all patterns at index $k$ if $\Vec{T}[k]$ is {\tt false}
    (in this case, pattern \(3\) at index \(0\)). $\square$
\end{example}

\subsubsection{Homomorphic Pattern Matching}
Given the set, $\mathcal{Q}'$, of reduced representative graph patterns from the previous step, 
we find their homomorphic matches in the input graph (cf. lines~\ref{fl-match}--\ref{fl-match-end} of 
Procedure~\ref{alg:patminer}). 
To ensure efficient matching of any pattern $Q\in\mathcal{Q}'$, we prune the input graph {\em w.r.t.}
the node and edge types in $Q$ to produced a simplified/filtered version of the input graph; 
then perform homomorphic sub-graph matching on the simplified graph with $Q$.

A {\it simplified graph}, $G'$, based on a given graph pattern $Q$ is a graph in 
which nodes and edges types (labels) that do not appear in $Q$ are removed 
\begin{wrapfigure}[8]{r}{0.4\textwidth}
\vspace{-12pt}
    \centering
    \includegraphics[]{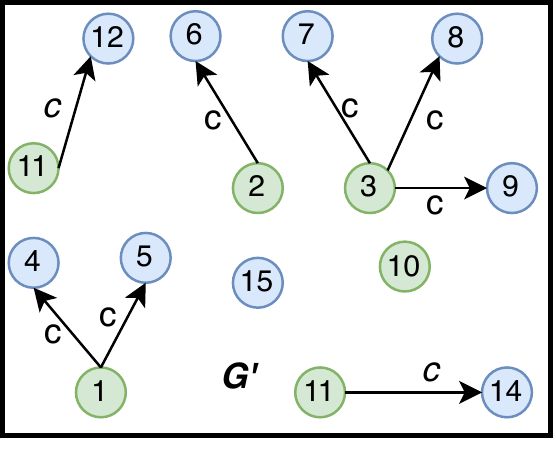}
    \caption{Simplified $G$, wrt $Q_3$}
    \label{fig:simplified-graph}
\end{wrapfigure}
from the original graph, $G$. 
The simplification of the input graph {\it w.r.t.} a graph pattern reduces the search space
of candidate matches significantly, and improves the efficiency of the matching process.
We adopt the efficient worst-case optimal join (WCOJ) based algorithm introduced in~\cite{b12} to
find all homomorphic matches of $Q$ in $G'$. 
For our running example, for brevity, suppose
we want to simplify sample graph $G$, {\em w.r.t.} pattern $Q_3[x,y,y']$ 
in Figure~\ref{fig:GED}~(a) and~(d), respectively. The resulting 
simplified graph, $G'$, is as shown in Figure~\ref{fig:simplified-graph}.

For each set of subgraph matched records based on a graph pattern, a large attributes’ relationship 
data table is generated by combining all attributes of entities in the graph pattern and the values 
of all attributes of entities in every matched subgraph. This pseudo-relation table stores the values 
of all attributes of entities in a graph pattern. The first row represents all attributes of entities 
contained in the graph pattern, and each of the remaining rows is information about a matching subgraph, 
representing the values of all attributes of entities in a matched subgraph.
An example of pseudo-relations over graph pattern $Q_3$ (from the running example in Figure~\ref{fig:GED})
is presented in Figure~\ref{fig:eg}.

\subsection{Attribute (Entity) Dependency Discovery}
Here, we mine dependencies over the pseudo-relational tables produced from the matches of 
the discovered graph patterns in the previous step. As depicted in Figure~\ref{fig:pipeLine}(B),
the dependency discovery architecture consists of four key modules, namely:
(a) Matches Preprocessor, (b) Partitioner, (c) Candidate Generator, and (d) the Validator. 

In the following, we discuss the function of each module (in order), illustrating the 
overall dependency mining algorithm over matches $H(Q, G)$ of a pattern $Q$ in a graph $G$ (as
captured in Procedure~\ref{pro:dep}).

\begin{procedure}
\caption{mineDependencies($Q,H(Q, G)$)}
\label{pro:dep}
\begin{algorithmic}[1] 
\Require A graph pattern $Q$, and its homomorphic matches $H(Q, G)$ in $G$
\Ensure A set $\Sigma$ of minimal GEDs over $Q$ in $G$
\State{$\Sigma := \emptyset; \mathbb{L}[0]:=\emptyset; N:=$ no. of unique attributes in $Q$}
\State{$\cal{T}\leftarrow $ form pseudo-relation of $H(Q, G)$ over $Q$}
\State{$\mathbb{L}[1] \leftarrow generateLiteralSetPartitions(Q[\bar{u}],\cal{T})$}
\State{${\sf LHS} \leftarrow \mathbb{L}[0]; {\sf RHS}\leftarrow \mathbb{L}[1]$}
\State{$\Sigma \leftarrow validateDependencies(\sf{LHS}, \sf{RHS})$}
\For{$i=2$ to $i=N-1$}
    \If{$|\mathbb{L}[i-1]|> 1$}
        \For{{\bf each} pair $(n_{1}, n_{2}) \in \mathbb{L}[i-1]\times \mathbb{L}[i-1]$}
            \State{$n_{} = n_1\cup n_2$}
            \If{$n$ is permissible}\Comment{Lemma~\ref{lm:permissible}}
                \State{$\mathbb{L}[i] := \mathbb{L}[i]\cup n$}
            \EndIf
        \EndFor
        \State{${\sf LHS}\leftarrow \mathbb{L}[i-1]$; ${\sf RHS}\leftarrow\mathbb{L}[i]$}
        \State{$\Sigma := \Sigma\cup validateDependencies({\sf LHS, RHS})$}
    \EndIf
\EndFor
\State{\Return $\Sigma$}
\end{algorithmic}
\end{procedure}


\subsubsection{Pre-processing matches of patterns}
This is an important (but optional) first step in the dependency mining process. In the 
presence of any domain-knowledge (e.g., in the form of meta data) about the input graph, 
one may leverage such knowledge to enhance the discovery process. For instance, given any 
graph pattern and its list of matches, based on available meta data, we can determine:
(a) which sets of attributes to consider for each pattern variable; 
(b) which pairs of pattern variables to consider for the generation of variable and $id$ 
literals.

That is, this phase allows any relevant pre-processing of the pseudo-relational
table of the matches of a graph pattern for efficient downstream mining tasks. 

\subsubsection{Partitioning}
Next, we transform the input table into structures for efficient mining of rules. In the following,
we extend the notions of {\em equivalence classes} and {\em partitions}~\cite{part} of 
attribute value pairs over pattern matches. This enables the adaption and use 
of relevant data structures, and efficient validation of candidate dependencies.  

\begin{definition}[Equivalence classes, partitions]\label{def:eq}
Let $H$ be the set of all matches of a graph pattern $Q[\bar{u}]$ in $G$, and 
$X$ be a set of literals in $\bar{u}$. 
The {\em partition} of $H$ by the set $X$ of literals, denoted by $\pi(X, H)$,
is the set of non-empty disjoint subsets, where each subset contains all matches
that satisfy $X$ with same value. For brevity, we write $\pi(X,H)$ as $\pi(X)$, 
when the context is clear.

Given $X, H$: $\pi(X,H) = \bigcap_{w\in X}\pi(w, H)$, where $w$~\footnote{for $x,y\in \bar{u}$,
$w$ can be: $(x.A= c)$, or $(x.A=y.A')$, or $(x.id=y.id)$} is a literal over pattern variable(s) in $Q[\bar{u}]$. 
Two sets of literals $X,X'$ are {\em equivalent  w.r.t.} $H$, iff: $\pi(X,H)=\pi(X',H)$. 
$\square$
\end{definition}

From the semantics of GED satisfaction and Definition~\ref{def:eq}, the following 
Lemma holds. 

\begin{lemma}\label{lm:valid}
A dependency $X\to Y$ holds over the pattern $Q[\bar{u}]$ in $G$ iff: $\pi(X,H)=\pi(X\cup Y, H)$.
\end{lemma}

\begin{figure}[!t]
\centering
\includegraphics[width=0.85\linewidth]{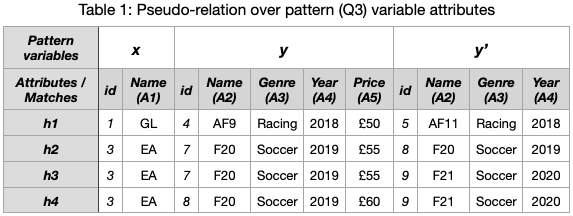}
\caption{Example of pseudo-relation over pattern $Q_3$ in Figure~\ref{fig:GED}(d)}
\label{fig:eg}
\end{figure}

\begin{figure}[!h]
\centering
\includegraphics[width=\linewidth]{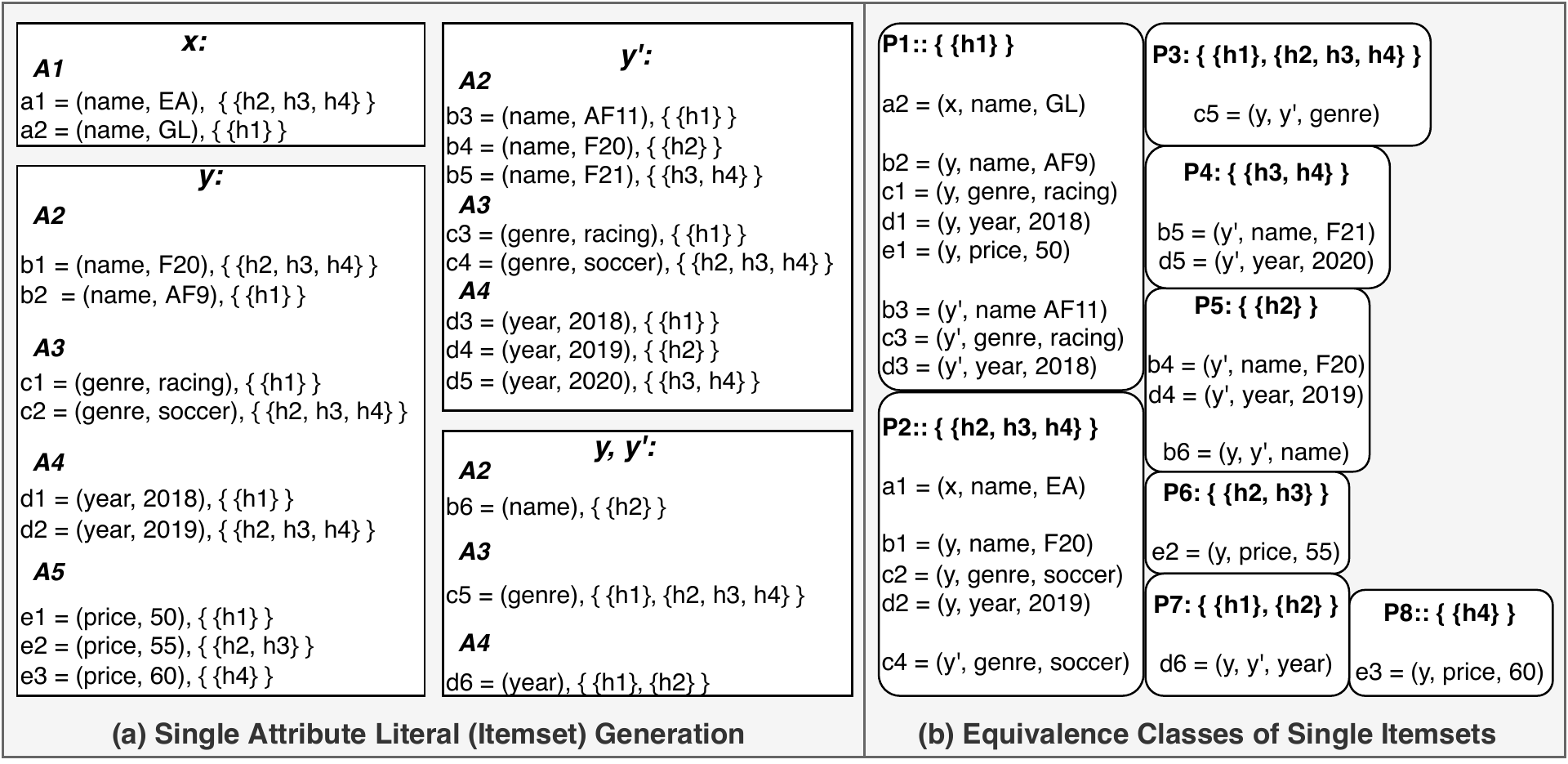}
\caption{Example of Pattern Variable Partitions \& Equivalence Classes.}
\label{fig:partitions}
\end{figure}

\begin{function}[!t]
\caption{$generateLiteralSetPartitions(Q[\bar{u}],\cal{T})$}
\label{func:part}
\begin{algorithmic}[1] 
\State{$\mathbb{L}[1]$ = $\emptyset$}
\For{{\bf each} pattern variable $x\in Q[\bar{u}]$}\Comment{constant literals}\label{fl-constant}
    \For{{\bf each} attribute $x.A \in \cal{T}$}
        \State{$L:= \{X=(c, x.A) \mid \forall \mbox{ } c\in adom(x.A) \wedge \pi(X, H(Q,G))\neq \emptyset\}$}
        \State{$\mathbb{L}[1]$.add($n$): $n= (X,\pi(X,H))$}
    \EndFor
\EndFor\label{fl-constant-end}
\For{{\bf each} pair $x,y\in Q[\bar{u}]$}\Comment{variable literals}\label{fl-variable}
    \State{$L:=\{X=(x.A=y.A')\mid A,A'\in {\cal T} \wedge \pi(X, H(Q,G))\neq \emptyset\}$}
    \State{$\mathbb{L}[1]$.add($n$): $n= (X,\pi(X,H))$}
\EndFor\label{fl-variable-end}
\State{\Return $\mathbb{L}[1]$}
\end{algorithmic}
\end{function}

In Function~\ref{func:part}, we presents the process of partitioning pattern variables
in $Q[\bar{u}]$ of a given pseudo-relational table $\cal{T}$.
Lines~\ref{fl-constant}--\ref{fl-constant-end} generates constant literals, whereas 
Lines~\ref{fl-variable}--\ref{fl-variable-end} of the function generates variable and $id$
literals. We remark that appropriate preprocessing (based on domain knowledge) from the 
previous step ensures only semantically meaningful pattern variable pairs are considered
for variable literals. 

In the following example, we show the partitions and equivalence classes of a
toy example. 

\begin{example}\label{ex:GED}
Suppose pseudo-relational in Figure~\ref{fig:eg} captures the attributes of the 
homomorphic matches of pattern $Q_3$ in Figure~\ref{fig:GED}. 
Figure~\ref{fig:partitions} shows the resulting partitions and equivalence classes.
We remark that, in this example, no $id$ literals are produced due to 
unique node identities in Figure~\ref{fig:eg}. 
\end{example}

\subsubsection{Candidate dependency generation}
Here, we discuss the data structures and strategies for generating candidate 
dependencies. We extend and model the search space of the possible dependencies with the
attribute lattice~\cite{lattice}. We adopt a level-wise top-down search approach for 
generating candidate rules, testing/validating them and using the valid rules to prune
the search space. 

\begin{figure}[!th]
\centering
\includegraphics[width=\textwidth]{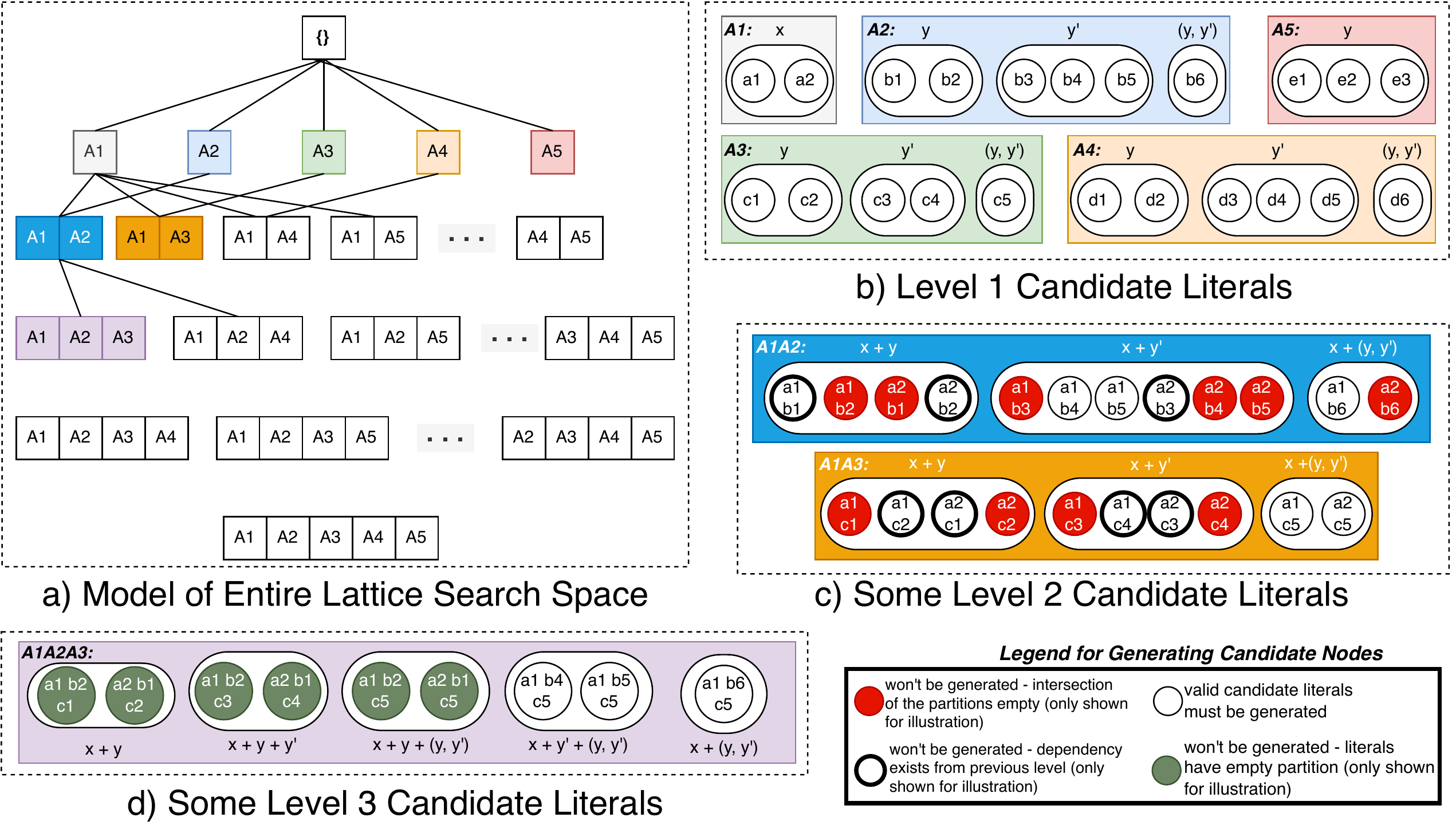}
\caption{GED Lattice: Generation, Search \& Traversal (best viewed in colour)}
\label{fig:lattiec}
\end{figure}  

In general, a node $n$ in the lattice is a pair $n = (X, \pi(X))$. However, 
for non-redundant generation and pruning of candidate sets $X, Y$, we organise 
literal sets of each node under their respective pattern variable(s). For example, Figure~\ref{fig:lattiec}~(a) shows a snapshot of the lattice for generating candidate 
dependencies (for the running example in Figure~\ref{fig:eg} and
Figure~\ref{fig:partitions}). 
Further, parts~(b), (c) and (d) of Figure~\ref{fig:lattiec} show all nodes in level 1, and some of the nodes for level 2 and 3 respectively.

To generate the fist level, $\mathbb{L}[1]$, nodes in the lattice, we use the single attribute
partition of variables in the previous step (i.e., output of Function~\ref{func:part}). 
Further, to derive nodes for higher levels of the lattice, we follow the following principles in
Lemma~\ref{lm:permissible} to ensure redundancy-free and valid literal set generation. 

\begin{lemma}[Permissible Node]\label{lm:permissible}
Let $\mathbb{L}[i]$ be the list of nodes at a current level $i$ of the lattice, with at least
two nodes $\mathbb{L}[i]=[n_1=(X_1, \pi(X_1)), n_2=(X_2, \pi(X_2))]$. We generate a new node 
$n = (X, \pi(X)) \in \mathbb{L}[i+1]$ using $n_1, n_2$, with the following properties:
(a) $X = X_1\cup X_2$; and (b) $\pi(X) = \pi(X_1)\cap \pi(X_2)$. 
We say $n$ is {\em permissible} iff: (i) $|X|> i$, (ii) $\nexists\mbox{ }w,w'\in X, \mbox{such that } 
x.A\in w,w'$ and (iii) $\pi(X)\neq \emptyset$. $\square$
\end{lemma}

For every node pair $n_1, n_2\in \mathbb{L}[i]$ that results in a {\em permissible} node 
$n\in \mathbb{L}[i+1]$, we establish the edges $n_1\to n$ and $n_2\to n$ representing 
candidate dependencies. All candidate dependencies are tested for validity in the next step. 
We illustrate the candidate dependency generation with our running example below. 

\begin{example}[Generating and searching the lattice space]\label{ex:candidate}
Figure~\ref{fig:lattiec} shows the general attribute lattice (left) and the extended literal set 
nodes associated with some of attribute sets (right). For brevity and clarity of presentation, 
we omit the partition set of nodes, and colour-code the non-permissible nodes. 

Candidate rules in $\mathbb{L}[0], \mathbb{L}[1]$ are of the form: $\emptyset\to Y$, 
where $Y$ is a single-attribute literal in $\mathbb{L}[1]$ (i.e., every level 1 node). The 
set of candidate rules in $\mathbb{L}[1], \mathbb{L}[2]$ over the edge
$A1\to A1A2$ take the form: $X\to Y$, where $X\in \{\{a1\}, \{a2\}\}$ and $Y
\in\{\{a1,b1\}, \{a1,b2\}, \cdots, \{a1,b6\}, \{a2,b6\}\}$. $\square$
\end{example}

\begin{function}[t]
\caption{$validateDependencies({\sf LHS, RHS})$}
\label{func:validate}
\begin{algorithmic}[1] 
\State{$\Sigma:=\emptyset$}
\For{{\bf each} node $n_i = (X_i, \pi(X_i))\in {\sf LHS}$}
    \For{{\bf each} node $n_j = (Y_j, \pi(Y_j)) \in {\sf RHS}$}
        \If{$X_i \subset Y_j$ and $\pi(X_i)=\pi(Y_j)$}\label{fl-edge}
            \State{$\sigma:= (Q[\bar{u}], X_i\to (Y_j\setminus X_i))$}\label{fl-valid}
            \State{$\Sigma:= \Sigma\cup \sigma$}
            \State{${\sf RHS}:={\sf RHS}\setminus n_j$}\label{fl-valid-end}
        \EndIf
    \EndFor
\EndFor
\State{\Return $\Sigma$}
\end{algorithmic}
\end{function}

\subsubsection{Validation of dependencies}
We use a level-wise traversal of search space to validate the generated candidate rules
between two successive levels in the lattice. That is, for any two levels $\mathbb{L}[i]$
and $\mathbb{L}[i+1]$, correspond to the set of {\sf LHS, RHS} candidates respectively. 
For any edge $X\to Y$ within {\sf LHS, RHS}, we test the dependency $X\to Y\setminus X$. 
Function~\ref{func:validate} performs this task, and its process is self-explanatory.
In Line~\ref{fl-edge}, $X_i\subset Y_j$ is true for all candidate dependencies (i.e., the
edge $X_i\to Y_j$ exists). 
If the dependency holds (i.e., $\pi(X_i)=\pi(Y_j)$ by Lemma~\ref{lm:valid}), we add the
GED to the set, and prune the node search space accordingly
(Lines~\ref{fl-valid}--\ref{fl-valid-end}). 

In Example~\ref{ex:validate}, we show the validation of some candidate GEDs in our
running example. 

\begin{example}[Validation of candidate GEDs]\label{ex:validate}
Consider the candidate GEDs in $\mathbb{L}[0],\mathbb{L}[1]$ in level 1 of the lattice,
from Example~\ref{ex:candidate}. 
It is easy to verify none of the candidates are valid, since the only literal set 
$X\in \mathbb{L}[0]$, has the property $X =(\emptyset, \pi(\emptyset, H))$. 
Note that $(\pi(\emptyset, H) =\{h1, h2, h3, h4\}$. More specifically, $\pi(X)\neq \pi(X\cup Y),
\forall\mbox{ } Y\in \mathbb{L}[1]$. 

Similarly, we can verify the GED $X\to Y$ holds in $\mathbb{L}[1], \mathbb{L}[2]$, 
where $X=(\{b1\}, \pi(\{b1\})), Y=(\{a1,b1\}, \pi(\{a1,b1\}))$, since 
$\pi(\{b1\}) =\pi(\{a1,b1\})$. Therefore, we can form the GED $(Q_3[x,y,y']$,
$y.name=\verb|F20|\to x.name=\verb|EA|)$, and prune the node $n=(\{a1,b1\}, 
\pi(\{a1,b1\}))\in\mathbb{L}[2]$. 
$\square$
\end{example}

\subsection{Minimal cover set discovery and ranking of GEDs}
In this phase of the GED mining process, we perform further
analysis of all mined rules in $\Sigma$ from the previous step to determine and 
eliminate all redundant or implied rules. Furthermore, we introduce a measuure
of interestingness to score and rank the resulting cover set $\Sigma_c$ of GEDs. 

\subsubsection{Minimal cover set}
From the foregoing, every GED $\varphi\in \Sigma$ is minimal (cf., Section~\ref{sec:pro}).
However, $\Sigma$ may not be minimal as there may exist redundant GEDs due to the
{\em transitivity} property of GEDs. More specifically, consider the following dependencies
in $\Sigma:= \{\varphi_1, \varphi_2, \varphi_3\}$; where $\varphi_1 = (Q[\bar{u}], X\to Y), 
\varphi_2 = (Q[\bar{u}],Y\to Z)$ and $\varphi_3 = (Q[\bar{u}], X\to Z)$. 
We say $\varphi_3$ is {\em transitively implied} in $\Sigma$, denoted by $\varphi_3\vDash \Sigma$ 
as $\Sigma\setminus \varphi_3 \equiv \Sigma$. 
That is, the set $\Sigma':= \{\varphi_1, \varphi_2\}\equiv \Sigma$, and $\Sigma'$ is minimal.

Thus, it suffices to eliminate all transitively implied GEDs in $\Sigma$ to 
produce a minimal cover $\Sigma_c$ of $\Sigma$. We use a simple, but effective 
process in Function~\ref{fun:minimal} to eliminate all transitively implied GEDs 
in $\Sigma$, and produce $\Sigma_c$. We create a graph $\Gamma$ with all GEDs in
$\Sigma$ such that each node corresponds to unique literal sets of the dependencies 
(i.e., Lines~\ref{fl-graph}, \ref{fl-graph-end}). 
We remark that, if there exists transitivity in $\Sigma$, triangles (aka. 3-cliques)
will be formed in $\Gamma$ (cf., Lines~\ref{fl-triangle}--\ref{fl-triangle-end}), and
a minimal set $\Sigma_c$ be returned (Lines~\ref{fl-minimal}--\ref{fl-minimal-end}). 

\begin{function}[t]
\caption{$findCover(\Sigma)$}
\label{fun:minimal}
\begin{algorithmic}[1] 
\State{$\Sigma_c:= \emptyset, \Gamma:=\emptyset$}
\For{{\bf each } $\varphi = (Q[\bar{u}], X\to Y) \in \Sigma$}\label{fl-graph}
\State{create the edge $e: X\to Y$ in $\Gamma$}\label{fl-graph-end}
\If{$e$ completes a triangle $t \in \Gamma$}\label{fl-triangle}
\State{analyse transitivity and prune $t$}\label{fl-triangle-end}
\EndIf
\EndFor
\For{{\bf each} remaining edge $e\in \Gamma$}\label{fl-minimal}
\State{add the corresponding GED of $e$ to $\Sigma_c$}
\EndFor
\State{\Return $\Sigma_c$}\label{fl-minimal-end}
\end{algorithmic}
\end{function}

\subsubsection{Ranking GEDs}
We present a measure to score the interestingness of GEDs as well as a means to
rank rules in a given minimal cover $\Sigma_c$. We adopt the minimum description length (MDL)
principle\cite{b20} to encode the relevance of GEDs. 
The MDL of a model $m$ is given by:
\[mdl(m, \mbox{data}) = cost(\mbox{data}|m) + \alpha\times cost(m),\]
where $cost(\mbox{data}|m)$, $cost(m)$ represent the cost of encoding errors by the model,
and cost of encoding the model itself, and $\alpha$ is a hyper-parameter for trade-off between
the two costs.

We propose in Equation~\ref{eq:mdl} below a measure for scoring the interestingness, $rank(\varphi, G)$,
of a GED $\varphi = (Q[\bar{u}], X\to Y)$ in a graph $G$ as:
\begin{equation}\label{eq:mdl}
rank(\varphi, G) = \alpha\times (1-\varsigma(\varphi)) + (1-\alpha)\times \frac{k}{N},
\end{equation}
where: 1) $\varsigma(\varphi)$ is the ratio $\frac{|\pi(X\cup Y)|}{|H|}\in [0,1]$, 
2) $k$ is the number of unique attributes in $X\cup Y \in [0,N]$, 
3) $N$ is the total number of attributes over all matches $H$ of $Q[\bar{u}]$, and
4) $\alpha \in [0,1]$ is a user-specified parameter to manipulate the trade-off between the two terms. 

In line with MDL, the lower the rank of a GED, the more interesting it is. We perform
an empirical evaluation of the score in the following section. 


\subsection{Complexity of Proposed Solution.}\label{ssec:xtime}
Here, we present sketch of the time complexity of our GED mining approach.
The key tasks in our solution are: the frequent subgraph mining (FSM), and the
attribute (entity) dependency mining (ADM) over the matches of the frequent subgraphs.
Thus, we present the theoretical and practical complexity of both tasks below.

\paragraph{FSM Complexity} Let $|G|$ and $k$ denote the number of nodes in 
a given graph and the largest graph pattern size respectively. Overall, in 
the worse case, the complexity of the FSM process is 
${\cal O}(2^{|G|^2}\cdot |G|^k)$ (cf. details in~\cite{b7}). This exponential 
time complexity {\em w.r.t.} $|G|$ is a major
performance bottleneck, as stated in Section~\ref{sec:intro}. Hence the graph
splitting heuristic strategy to reduce $|G|$ via dense community detection (cf.
subsection~\ref{ssec:fsm}).
Thus, in practice, our FSM time is ${\cal O}(C\cdot 2^{|N|^2}\cdot |N|^k)\approx 
{\cal O}(2^{|N|^2}\cdot |N|^k)$, 
where $C<<|N|<<|G|$ is the largest community size and $C$ is the total number 
of communities.

Further, performing a homomorphic matching from a $k$-sized subgraph to $G$,
can be done, in the worse case,  in ${\cal O}(|G|^2\cdot k)$ time (cf.~\cite{b12}
and~\cite{fsm}). Thus, overall time complexity for FSM is roughly ${\cal O}
(2^{|N|^2}\cdot |N|^k + |G|^2)$.

\paragraph{ADM Complexity}
Here we examine the complexity of searching and validating all dependencies
over the graph patterns. Let $|\bar{u}|$, $m$, $|H|$ be the number of
nodes in a given graph pattern $Q[\bar{u}]$, maximum number of attributes
per pattern variable $x\in \bar{u}$, and the total number of matches of
$Q[\bar{u}]$ in $G$.

Then, the complexity of generating partitions of $H$ by literals of $\bar{u}$ 
is ${\cal O}(|\bar{u}|^2\cdot m\cdot |H|)\approx {\cal O}(m|H|)$; since in practice, $|\bar{u}|
<<m<<|H|$ (cf. Function~\ref{func:part}). Moreover, the search space of 
candidate GEDs is $2^{m\cdot |\bar{u}|^2}$.
Thus, to generate, search \& validate, and reduce candidate GEDs over $Q[\bar{u}]$ in $H$, requires
${\cal O}(|\bar{u}|^2m|H|+2^{m|\bar{u}|^2}|H| + |\Sigma| + |E(\Gamma)|)\approx {\cal O}(mH+2^{m}|H|+|\Sigma|)$ in the 
worse case (cf. Functions~\ref{func:validate} and~\ref{fun:minimal}), where
$|\Sigma|$ is the number of valid rules, and $|E(\Gamma)|\leq |\Sigma|$ is the 
total edges in the rules graph $\Gamma$.

\section{Empirical Study}\label{sec:exp}
In this section, we present an evaluation of the proposed discovery approach using real-world
graph data sets. In the following, we discuss the setting of the experiments, examine the 
scalability of the proposed algorithm, compared our solution to relevant related works in the 
literature, and demonstrate the usefulness of some of the mined GEDs in various downstream applications.

\setcounter{table}{1}
\begin{table}[]
    \centering
    \caption{Summary of the data sets}
    \label{tab:data}
    \footnotesize
    \begin{tabular}{|l||c|c|c|c|c|}
    \hline
      {\bf Data Set}   & {\bf \#Nodes} & {\bf \#Edges} & {\bf \#Node Types} & {\bf \#Edge Types} & 
      {\bf Density}\\
      \hline\hline
      {\sf DBLP} & 300K & 800K & 3 & 3 & $8.65\times 10^{-6}$\\
      \hline
      {\sf IMDB} & 300K & 1.3M & 5 & 8 & $1.25\times 10^{-5}$\\
      \hline
      {\sf YAGO4} & 4.37M & 13.3M & 7764 & 83 & $8.70\times 10^{-6}$\\
      \hline
    \end{tabular}
\end{table}

\subsection{Experimental Setting}
Here, we cover the data sets used and the setting of experiments. 

{\bf Data Sets.} 
We used three real-world benchmark data sets with different features and sizes, 
summarised in Table~\ref{tab:data}. The {\sf DBLP}~\cite{b14} is a citation network 
with 0.3M nodes of 3 types and 0.8M edges of 3 types; {\sf IMDB}~\cite{b15} is a 
knowledge graph with 0.3M nodes of 5 types and 1.3M edges of 8 types; and 
{\sf YAGO4}~\cite{b16} is a knowledge graph with 4.37M nodes with 7764 types and 13.3M 
edges with 83 types. 
In line with~\cite{gar}, we use the top five most frequent attribute values in the
active domain of each attribute to construct constant literals.

{\bf Experiments.} 
All the proposed algorithms in this work were implemented in Java; and the
experiments were conducted on a 2.20GHz Intel Xeon processor computer with 128GB of 
memory running Linux OS. We experimentally evaluated our proposed method to show its: 
(1) the scalability, (2) relative performance {\em w.r.t.} mining other (related) graph 
constraints/dependencies in the literature, 
(3) the potential usefulness of the mined GEDs. 
Unless otherwise specified, we fix the resolution parameter $\gamma$ to $1.0$, and
the minimum MNI threshold $\tau$ to \(5000\) in all experiments.
All experiments were conducted twenty-five (\(25\)) times, and report the average of all results.


\subsection{Scalability of proposal}\label{sec:scale}
Based on the complexity analysis of our proposition in Section~\ref{ssec:xtime},
the key parameters that impact the time performance include: the size of the
input graph $|G|$, size of graph patterns $k$, and number of attributes $m$
per pattern variable.
Thus, in this set of experiments, we investigate
the time performance of the proposed solution on all three data sets with respect 
to varying: a) input graph sizes $|G|$ (Exp-1a), b) graph pattern sizes $k$ (Exp-1b), and 
c) number of attributes $m$ on nodes (Exp-1c). 

\begin{figure}[!th]
    \centering
    \includegraphics[width=\textwidth]{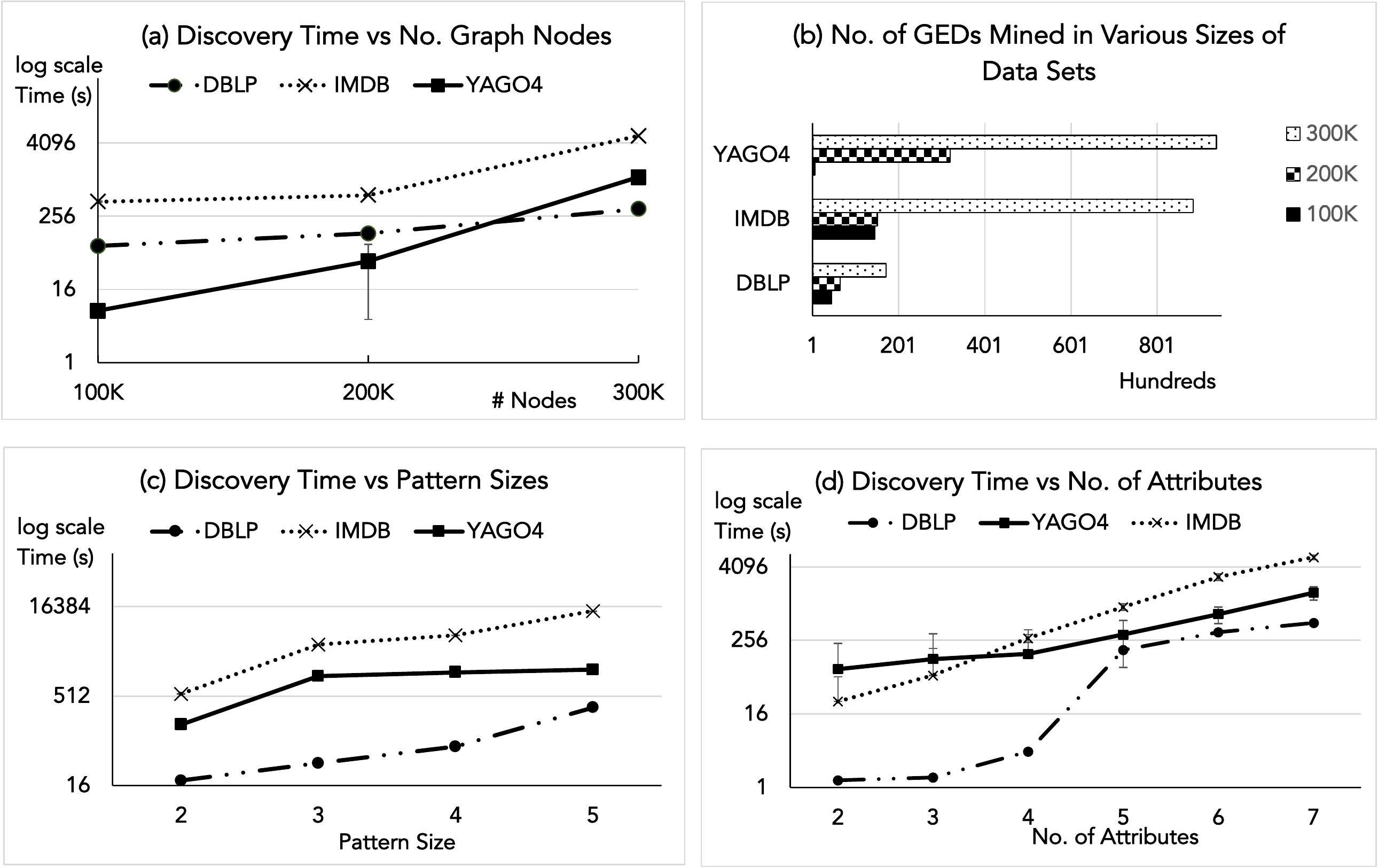}
    \caption{Scalability of proposed approach}
    \label{fig:scale}
\end{figure}

{\bf Exp-1a.} In this experiment, we fix the number of attributes on every node to 
seven (7) in all three data sets. We used the full graph of {\sf DBLP}, sampled 
comparable sizes of {\sf IMDB} and {\sf YAGO4} graphs. Figure~\ref{fig:scale}~(a)
presents a plot of the discovery time (in seconds, on a log-scale) against 
different sizes of the three data sets, with their corresponding standard 
deviations (SD) in Table~\ref{tab:sd-6a}. 

\begin{table}[!th]
\footnotesize
    \centering
    \begin{minipage}{0.45\textwidth}
    \caption{SD of Times in Figure~\ref{fig:scale}~(a)}
            \begin{tabular}{c|c|c|c}\hline
         & 100K & 200K & 300K \\ \hline
       {\sf DBLP} & \(2.927\) & \(5.266\) & \(7.236\)\\ \hline
       {\sf YAGO4}  & \(0.095\) & \(0.413\) & \(6.741\) \\ \hline
       {\sf IMDB} & \(4.491\) & \(0.749\) & \(23.066\)
    \end{tabular}
    \label{tab:sd-6a}
    \end{minipage}
    \begin{minipage}{0.45\textwidth}
    \caption{SD of Times in Figure~\ref{fig:scale}~(c)}
        \begin{tabular}{c|c|c|c|c}\hline
         & 2 & 3 & 4 & 5 \\ \hline
       {\sf DBLP} & \(0.09\) &\(2.16\) &\(1.90\) &\(2.67\) \\ \hline
       {\sf YAGO4}  &\(5.12\) &\(11.94\) &\(10.76\) &\(19.0\) \\ \hline
       {\sf IMDB} &\(7.06\) &\(14.30\) &\(23.07\) &\(20.95\)
        \end{tabular}
        \label{tab:sd-6c}
    \end{minipage}
\end{table}

In general, GED discovery in the {\sf DBLP} data is the most efficient as it produces the
least number of matches for its frequent patterns compared to the other data sets. This 
characteristics is captured by the density of the graphs (in Table~\ref{tab:data}). In other
words, the more dense/connected a graph is, the more likely it is to find more matches for graph
patterns. Consequently, the longer the GED discovery takes. 

The distribution of discovered GEDs, for different sizes of the three data sets is
presented in Figure~\ref{fig:scale}~(b) -- with the {\sf YAGO4} and {\sf DBLP} data 
sets producing the most and least number of GEDs in almost all cases respectively.

{\bf Exp-1b.} Here, we examine the impact of graph pattern sizes on the discovery time.
We use the number of nodes in a graph pattern as its size -- considering patterns of size
2 to 5. For each data set, we mine GEDs over patterns size 2 to 5, using the full graph with
up to seven (7) attributes per node.

The result of the time performance for different sizes of patterns are presented in 
Figure~\ref{fig:scale}~(c), with corresponding standard deviations in Table~\ref{tab:sd-6c}. 
In this experiment, the {\sf IMDB} and {\sf YAGO4} data sets
produced comparable time performances, except for the case of patterns with size 5. 
In the exceptional case, we found significantly more matches of size-5 patterns in
the {\sf IMDB} data, which is reflected in the plot.

{\bf Exp-1c.} This set of experiments examine the impact of the number 
of attributes on the discovery time in all data sets, varying the number of attributes on each node in
the graphs from 2 to 7. The time performance of graph patterns of all sizes on the three datasets is averaged as the final result. 

The plots in Figure~\ref{fig:scale}~(d) show the discovery average time characteristics 
\begin{wraptable}[5]{r}{0.57\textwidth}
\vspace{-20pt}
\footnotesize
    \centering
    \caption{SD of Times in Figure~\ref{fig:scale}~(d)}
        \begin{tabular}{c|c|c|c|c|c|c}\hline
         & 2 & 3 & 4 & 5 & 6 & 7\\ \hline
       {\sf DBLP} & \(0.04\) &\(0.02\) &\(0.03\) &\(0.07\) &\(0.04\) &\(0.27\) \\ \hline
       {\sf YAGO4}  &\(0.02\) &\(0.07\) &\(0.06\) &\(0.13\) &\(0.27\)&\(0.60\)\\ \hline
       {\sf IMDB} &\(0.05\) &\(0.12\) &\(0.88\) &\(2.76\) & \(2.03\)& \(2.32\)
        \end{tabular}
        \label{tab:sd-6d}
\end{wraptable}
for all three data sets with increasing number of attributes (for 
patterns with sizes from \(2\) to \(5\)),
with standard deviations in Table~\ref{tab:sd-6d}. 
As can be seen from all three line graphs, the number of attributes directly affects 
the time performance, as expected by the analysis of
the characteristics (cf. Section~\ref{ssec:xtime}).

\subsection{Comparison to mining other graph constraints}
The notion of GEDs generalises and sunbsumes graph functional dependencies (GFDs) 
and graph keys (GKeys).
Thus, we show in this group of experiments that our proposal can be used for mining 
GFDs and GKeys via two simple adaptions. 
First, we replace the graph pattern matching semantics from homomorphism
to isomorphism in accordance with the definitions of GFDs and GKeys. 
Second, we consider only id-literal RHSs for GKey mining, 
and no id-literals in GFD mining.

\begin{figure}[htbp]
\centering
\includegraphics[width=\linewidth]{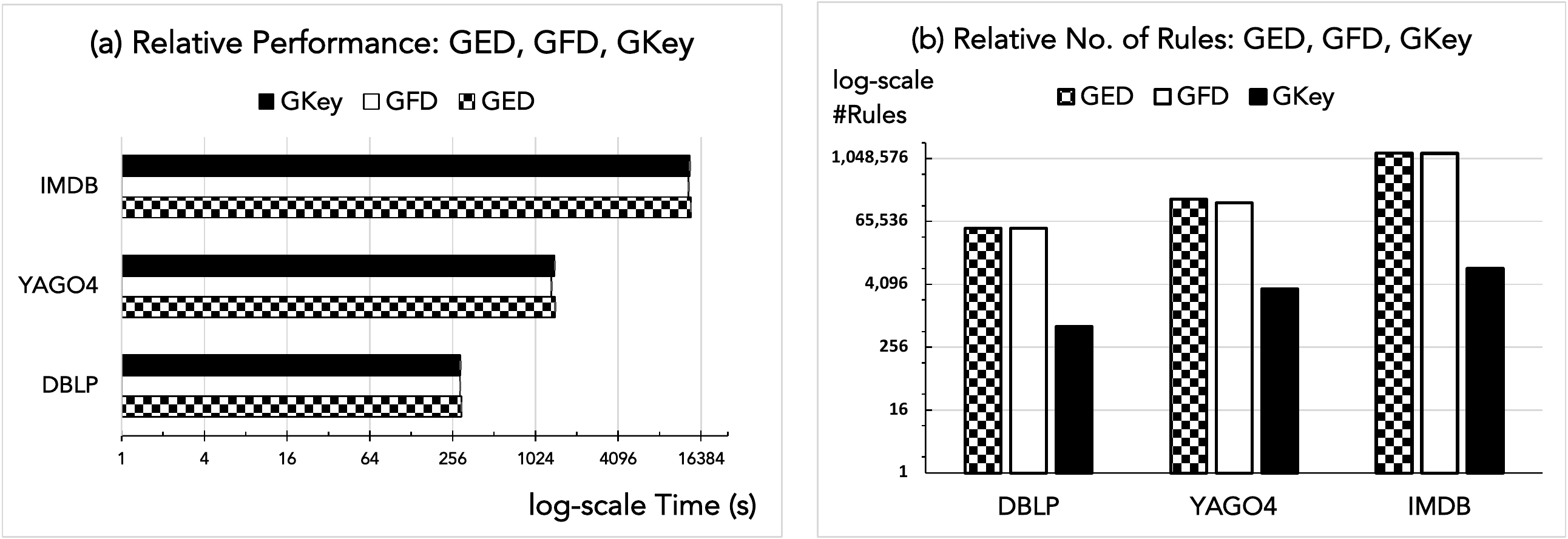}
\caption{The performance comparison of mining GED, GFD and Gkey}
\label{fig:comp}
\end{figure}

In Figure~\ref{fig:comp}, we present a summary of the comparative analysis of mining GEDs, 
GFDs and GKeys. Parts~(a) of the plot represents the relative time performance of
mining the three different graph constraints over the {\sf DBLP, YAGO4} and {\sf IMDB} 
data sets respectively. In each experiment, we show the performance for up to the five (5) 
graph patterns (using the top five most frequent patterns in each data set).

\begin{wraptable}[5]{r}{0.43\textwidth}
\footnotesize
\vspace{-20pt}
  \caption{SD of Times in Figure~\ref{fig:comp}~(a)}
            \begin{tabular}{c|c|c|c}\hline
         & GED & GFD & GKey \\ \hline
       {\sf DBLP} & \(0.71\) & \(0.35\) & \(0.60\)\\ \hline
       {\sf YAGO4}  & \(1.92\) & \(0.86\) & \(0.87\) \\ \hline
       {\sf IMDB} & \(15.67\) & \(14.85\) & \(16.20\)
    \end{tabular}
    \label{tab:sd-7a}
\end{wraptable}
As expected, the differences in the time performances are minimal in all cases, as
we have optimised: a) the matching efficiency of homomorphism vs. isomorphism; 
and b) the number of literals considered for each dependency/constraint type. 
Table~\ref{tab:sd-7a} presents the standard deviation of the time performance on
all datasets.

Furthermore, we show in Figure~\ref{fig:comp}~(b) the number of discoveries 
for each constraint over all datasets. In particular, the number of GEDs versus 
GFDs is consistently similar in all datasets; and GKeys are the smallest sized
rules as the RHSs are restricted to onlyid-literals.

\subsection{Usefulness of mined rules}
In this section, we present some examples of the discovered GEDs with a discussion of their
potential use in real-world data quality and data management applications. Further, we discuss
interestingness ranking of the mined GEDs.

\begin{figure}[tbp]
\centering
\includegraphics[width=0.95\linewidth]{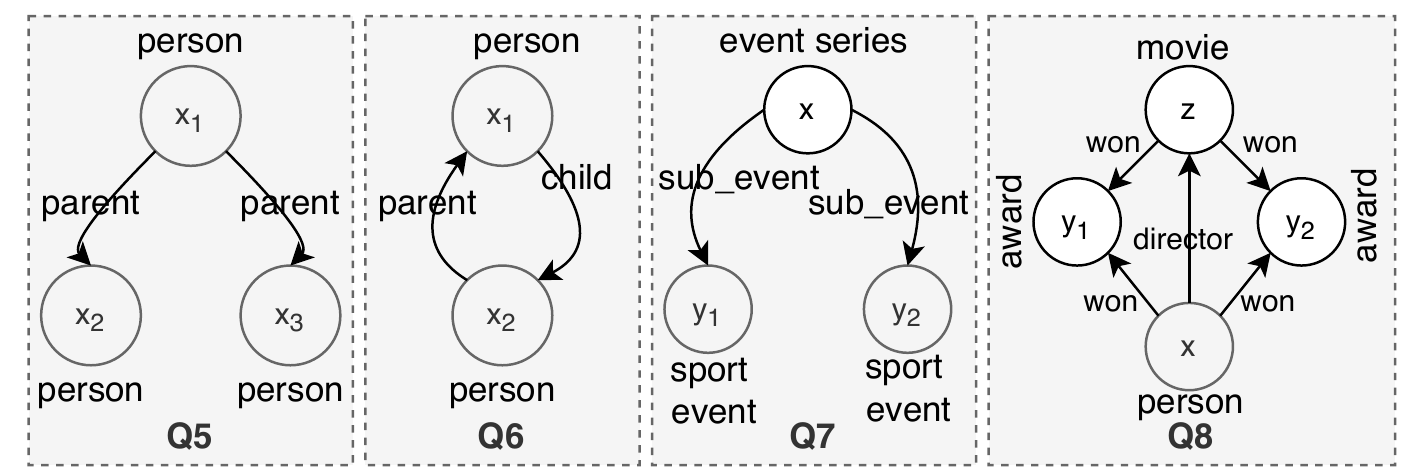}
\caption{Graph patterns for some GEDs in {\sf YAGO4}}
\label{meaningful_GEDs}
\end{figure}

{\bf Examples of discovered GEDs.}
We manually inspected the mined GEDs and validated their usefulness. We present 3 GEDs found in the {\sf YAGO4} data, with patterns shown in Figure~\ref{meaningful_GEDs} as follows:

\begin{itemize}
    \item[$\sigma_1:$] $Q_5[x_1,x_2,x_3], \{x_1.{\tt dob} = \verb|1777-04-24|,x_2.{\tt dod}=\verb|1792-03-01|,x_3.{\tt name}=\verb|`Leopold II, Holy Roman Emperor'|\}\to x_2.{\tt id}=x_3.{\tt id}$.
    \item[$\sigma_2:$] $Q_6[x_1, x_2], \emptyset \to x_1.{\tt surname}=x_2.{\tt surname}$.
    \item[$\sigma_3:$] $Q_{7}[x,y_1,y_2], x.{\tt name} =\verb|`2018 Wuhan Open'| \to x_1.{\tt prefixname}=x_2.{\tt prefixname}$.
    \item[$\sigma_4:$] $Q_8[x,y_1,y_2,z], x.{\tt name}=\verb|`Akira Kurosawa'| \wedge y_1.{\tt name}=\verb|`Golden Lion'| \wedge y_2.{\tt url}=\verb|`http:www.labiennale.org'| \wedge z.{\tt name}=\verb|`Rashomon'| \to y_1.{\tt id} =y_2.{\tt id}$
\end{itemize}

\sloppy
The GED $\sigma_1$ states that for any match of the pattern $Q_5$ in the {\sf YAGO4} graph, if
the \verb|person| $x_1$ is born on \verb|1777-04-24|, and has a parent $x_2$ whose date of death is \verb|1792-03-01| with another parent $x_3$ named \verb|`Leopard II, Holy Roman Emperor'|, then
 $x_2,x_3$ refer to the same person. 
 This GED is verifiable true (see the Wikipedia record here\footnote{\url{https://en.wikipedia.org/wiki/Archduchess_Maria_Clementina_of_Austria}}) -- as person $x_1$ is the \verb|`Archduchess Maria Clementina of Austria'| a child of 
 \verb|`Leopard II, Holy Roman  Emperor'|. 
 The discovery of $\sigma_1$ in the {\sf YAGO4} data signifies duplicate \verb|person| entities 
for \verb|`Leopard II, Holy Roman Emperor'|.  Thus, a potential use for this GED can be for the
disambiguation or de-duplication (entity linkage) of said entities in the graph. 

Furthermore, $\sigma_2$ states that for any matches of the parent-child pattern $Q_6$, 
the child $x_1$ and parent $x_2$ must share the same last name. 
The GED $\sigma_3$ claims that for any two sport events $y_1,y_2$ in the \verb|2018 Wuhan Open| event series $x$; $y_1,y_2$ must share  the same event \verb|prefixName|. 
Indeed, these GED are straightforward and understandable; and can be used for violation
or inconsistency detection in the graph through the validation of the dependencies. 
For example, $\sigma_2$ can be used to check all 
parent-child relationships that do not have the same surname, whiles $\sigma_3$ can 
be employed to find any sub-event in the \verb|`2018 Wuhan Open'| that violate the
prefix-naming constraints. 
The GED $\sigma_4$, like $\sigma_1$, is an example of a GKey (i.e., have RHS id-literals). Thus, useful for entity resolution or de-duplication.

\begin{figure}[!t]
\centering
\includegraphics[width=\linewidth]{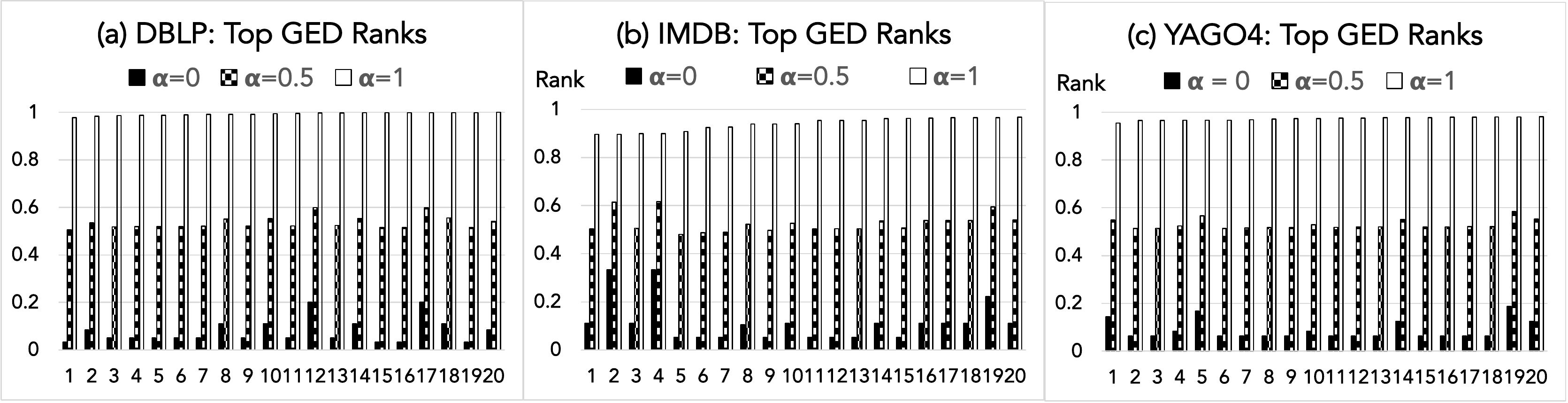}
\caption{Top ranked GEDs w.r.t. $\alpha$ settings}
\label{exp5}
\end{figure}

{\bf The rank of GEDs.}
In the following, we show a plot of the interestingness scores of the discovered GEDs, for three different $\alpha$ values (i.e. 0,0.5 and 1, respectively). 
For brevity, and lack of space, we show the results for only the top $K=20$ GEDs in
Figure \ref{exp5}. A lower interestingness score is representative of better GED, and vice versa.
The plot shows the interestingness score is effected by the value of $\alpha$, and ranking GEDs makes it easier for users to find meaningful GEDs.
For $\alpha=0$, the rank score reflects the complexity of the GED, whereas $\alpha=1$
reflects the persistence of the rule in data. And, $\alpha=0.5$ combines both with equal 
weight.

\section{Related Work}\label{sec:rw}
In this section, we review related works in the literature on the discovery of graph data
constraints. 
Mining constraints in graph data has received increasing attention in recent years.
In the following, we present some relevant related works in two broad categories.

\subsection{Rule discovery in non-property graphs.}
Most research on profiling and mining of non-property graph data focus on
XML and RDF data (cf.~\cite{data_profiling}). For instance, \cite{sem_rdf,ar_rdf,path_ar}
investigate the problem of mining (path) association rules in RDF data and knowledge 
bases (KBs), whereas \cite{amie,scale_kb,amie3,rudik} present inductive logic programming 
based rule miners over RDF data and KBs. 

In contrast to the above-mentioned works, this paper studies the mining of functional 
(both conditional and entity) dependencies in property graphs. However,
our proposed method can be adapted for mining rules in RDF data, particularly,
GEDs with constant literals. 

\subsection{Rule discovery in property graphs.}
More related to this work, are techniques for rule discovery in property graphs. 
Examples of some notable works in this area include: 1)~\cite{b4,gar,gtar,carg,targ}, which 
investigated the discovery of association rules in property graphs; and 2)~\cite{gkeyminer, 
b3, b1, tgfd} on mining keys and dependencies in property graphs---closest to this work. 
In particular, \cite{gkeyminer} presents a frequent sub-graph expansion based approach
for mining keys in property RDFs, whiles~\cite{b3} proposes efficient 
sequential and parallel graph functional dependency discovery for large property graphs. 
The main distinction between the GFD discovery approach in~\cite{b3} and our work is 
on how the efficiency bottlenecks are handled. Specifically,~\cite{b3} employs 
fixed parameterization to bound the sized of patterns to be mined, whereas we
modulate the graph and mine graph patterns from dense communities in the graph.
Furthermore, we devise a level-wise search strategy to find and validate rules
via an itemset lattice,
while the sequential technique in~\cite{b3} finds GFDs via a vertical and 
horizontal spawning of generation trees.

Moreover, the work in~\cite{tgfd} studies the discovery of {\it temporal} GFDs -- 
GFDs that hold over property graphs over periods of time; and~\cite{b1} studies
the discovery of graph {\it differential} dependencies (GDDs)---{\it GEDs with distance
semantics}---over property graphs.
Although related, this work differs from mining temporal GFDs as we consider only one 
time period of graphs. And, we do not consider the semantics of difference in data as 
captured in GDDs. Essentially, GEDs can be considered a special case of GDDs, where
the distance is zero over all attributes.

\section{Conclusion}
In this paper, we presented a new approach for mining GEDs. The developed discovery pipeline seamlessly combines graph
partitioning, non-redundant and frequent graph pattern mining, homomorphic graph pattern matching, and attribute/entity dependencies mining to discover GEDs in property graphs. We develop effective 
pruning strategies and techniques to ensure the returned set of GEDs is minimal, without redundancies. 
Furthermore, we propose an effective MDL-based measure to score the interestingness of GEDs. 
Finally, we performed experiments on large real-world graphs, to demonstrate the feasibility,
effectiveness and scalability of our proposal. 
Indeed, the empirical results are show our method is effective, salable and efficient; and finds semantically meaningful rules.

\section*{Acknowledgment}
This work was supported in part by Inner Mongolia Key Scientific and Technological Project under Grant 2021ZD0046, and in part by Innovation fund of Marine Defense Technology Innovation Center under Grant JJ-2021-722-04, and in part by the Major Project of Hubei Hongshan Laboratory under Grant 2022HSZD031, and in part by the open funds of the National Key Laboratory of Crop Genetic Improvement, Huzhong Agricultural University, and in part by the open funds of the State Key Laboratory of Agricultural Microbiology, Huzhong Agricultural University, and in part by 2021 Open Project of State Key Laboratory of Hybrid Rice, Wuhan University, and in part by the Fundamental Research Funds for the Chinese Central Universities under Grant 2662020XXQD01 and 2662022JC004.

\end{document}